\documentclass[useAMS,usenatbib]{mn2e}
\usepackage{epsfig,amsmath,amssymb,amsfonts,mathrsfs,latexsym,graphicx}

\usepackage[draft]{hyperref}
\usepackage{natbib}
\usepackage{color}
\usepackage{algorithm}
\usepackage{ textcomp } 

\usepackage{fixltx2e}

\usepackage{float}
\usepackage[caption = false]{subfig}
\usepackage{multirow}
\include{journals}

\graphicspath{{images/}}

\newcommand{\Msun}{{\rm M}_{\odot}}

\newcommand{\kms}{km~s$^{-1}$}

\newcommand{\SII}{S~{\sc ii}}

\newcommand{\SiII}{Si~{\sc ii}}

\newcommand{\CaII}{Ca~{\sc ii}}

\newcommand{\Deltam}{$\Delta m_{15}(B)$}

\def\gsim{\mathrel{\rlap{\lower 4pt \hbox{\hskip 1pt $\sim$}}\raise 1pt \hbox {$>$}}}
\def\lsim{\mathrel{\rlap{\lower 4pt \hbox{\hskip 1pt $\sim$}}\raise 1pt \hbox {$<$}}}

\title[A metric space for SNe Ia]{ A metric space for type Ia supernova spectra}

\author[M. Sasdelli et al.]
{\parbox{\textwidth}{\vspace{-.5cm} \large  Michele~Sasdelli$^{1}$\thanks{E-mail: 
sasdelli@mpa-garching.mpg.de}, 
 W.~Hillebrandt$^{1}$,
G.~Aldering$^{2}$,
P.~Antilogus$^{3}$,
C.~Aragon$^{2}$,
S.~Bailey$^{2}$,
C.~Baltay$^{4}$,
S.~Benitez-Herrera$^{1}$,
S.~Bongard$^{3}$,
C.~Buton$^{5,6}$,
A.~Canto$^{3}$,
F.~Cellier-Holzem$^{3}$,
J.~Chen$^{7}$,
M.~Childress$^{2,8}$,
N.~Chotard$^{7,9}$,
Y.~Copin$^{10}$,
H.~K.~Fakhouri$^{2,8}$,
U.~Feindt$^{5,6}$,
M.~Fink$^{1}$,
M.~Fleury$^{3}$,
D.~Fouchez$^{11}$,
E.~Gangler$^{10}$,
J.~Guy$^{3}$,
E.~E.~O.~Ishida$^{1,12}$,
A.~G.~Kim$^{2}$,
M.~Kowalski$^{5,6}$,
M.~Kromer$^{1,13}$,
S.~Lombardo$^{5,6}$,
 P.~A.~Mazzali$^{14,1,15}$,
J.~Nordin$^{2,16}$,
R.~Pain$^{3}$,
E.~P\'econtal$^{18}$,
R.~Pereira$^{10}$,
S.~Perlmutter$^{2,8}$,
D.~Rabinowitz$^{4}$,
M.~Rigault$^{5,6}$,
K.~Runge$^{2}$,
C.~Saunders$^{2}$,
R.~Scalzo$^{19}$,
G.~Smadja$^{10}$,
N.~Suzuki$^{2}$, 
C.~Tao$^{7,11}$,
S.~Taubenberger$^{1}$,
R.~C.~Thomas$^{17}$,
A.~Tilquin$^{11}$,
B.~A.~Weaver$^{20}$
}\vspace{0.6cm}\\
\parbox{\textwidth}{ \tiny
$^{1}$ Max-Planck-Institut f\"ur Astrophysik, Karl-Schwarzschild-Str. 1, 85741 Garching bei M\"unchen, Germany\\
$^{2}$Physics Division, Lawrence Berkeley National Laboratory,
  1 Cyclotron Road, Berkeley, CA 94720, USA\\
$^{3}$   Laboratoire de Physique Nucl\'eaire et des Hautes \'Energies, Universit\'e Pierre et Marie Curie Paris 6, Universit\'e Paris Diderot Paris 7,
 CNRS-IN2P3, 4 place Jussieu, 75252 Paris Cedex 05, France\\
$^{4}$Department of Physics, Yale University,
  New Haven, CT 06520-8121, USA\\
$^{5}$Physikalisches Institut, Universit\"at Bonn,
  Nu\ss allee 12, 53115 Bonn, Germany\\
$^{6}$Institut für Physik, Newtonstr. 15, 12489 Berlin,
 Humboldt-Universität zu Berlin\\
$^{7}$Tsinghua Center for Astrophysics, Tsinghua University, Beijing
  100084, China\\
$^{8}$Department of Physics, University of California Berkeley,
  366 LeConte Hall MC 7300, Berkeley, CA, 94720-7300, USA\\
$^{9}$National Astronomical Observatories, Chinese Academy of Sciences,
  Beijing 100012, China\\
$^{10}$Universit\'e de Lyon, 69622, France; Universit\'e de Lyon 1, France;
  CNRS/IN2P3, Institut de Physique Nucl\'eaire de Lyon,
  France\\
$^{11}$Centre de Physique des Particules de Marseille, 
Aix Marseille Universit\'e, CNRS/IN2P3, CPPM UMR 7346, 13288 Marseille, France\\
$^{12}$IAG, Universidade de S\~ao Paulo, Rua do Mat\~ao 1226, Cidade Universit\'aria,
CEP 05508-900, S\~ao Paulo, SP, Brazil\\
$^{13}$The Oskar Klein Centre \& Dept. of Astronomy, Stockholm University,
  AlbaNova, SE-106 91 Stockholm, Sweden\\
$^{14}$Astrophysics Research Institute, Liverpool John Moores University,
Liverpool L3 5RF, UK \\
$^{15}$INAF-Osservatorio Astronomico di Padova, vicolo dell'Osservatorio, 5, I-35122 Padova, Italy\\
$^{16}$Space Sciences Laboratory, University of California Berkeley, 7 Gauss 
Way, Berkeley, CA
  94720, USA\\
$^{17}$Computational Cosmology Center, Computational Research Division,
  Lawrence Berkeley National Laboratory, 1 Cyclotron Road
  MS~50B-4206,
  Berkeley, CA, 94720, USA\\
$^{18}$Centre de Recherche Astronomique de Lyon, Universit\'e Lyon 1,
  9 Avenue Charles Andr\'e, 69561 Saint Genis Laval Cedex, 
France\\
  $^{19}$Research School of Astronomy and Astrophysics,
    The Australian National University,
    Mount Stromlo Observatory,
    Cotter Road, Weston Creek ACT 2611 Australia\\
  $^{20}$ Center for Cosmology and Particle Physics,
    New York University,
    4 Washington Place, New York, NY 10003, USA\\
}}

\begin{document}

\date{Accepted ... Received ...; in original form ...}

\pagerange{\pageref{firstpage}--\pageref{lastpage}} \pubyear{2014}

\maketitle
\label{firstpage}

\begin{abstract}
We develop a new framework for use in exploring Type Ia Supernova (SN~Ia) spectra. Combining Principal Component Analysis (PCA) and Partial Least Square analysis (PLS) we are able to establish correlations between the Principal Components (PCs) and spectroscopic/photometric SNe~Ia features.  The technique was applied to $\sim120$ supernova and $\sim800$ spectra from the Nearby Supernova Factory. The ability of PCA to group together 
SNe~Ia with similar spectral features, already explored in previous studies, is greatly enhanced by two important modifications: (1) the initial data matrix is built using derivatives of spectra over the wavelength, which increases the weight of weak lines and discards extinction, and (2) we extract time evolution information  through the use of entire spectral sequences concatenated  in each line of the {input} data matrix.  These allow us to  define a stable PC parameter space which can be used to characterize synthetic SN~Ia spectra by means of real SN features.
 Using PLS, we  demonstrate  that the information from important previously known spectral indicators (namely the pseudo-equivalent width (pEW) of \SiII~5972~\AA/\SiII~6355~\AA\ and the line velocity of \SII~5640~\AA/\SiII~6355~\AA) at a given epoch, is contained within the PC space and can be determined through a linear combination of the most important PCs.  We also show that the PC space encompasses photometric features like  B/V magnitudes, B-V colors and SALT2 parameters $c$ and $x_1$. 
{The observed colors and magnitudes, that are heavily affected by extinction, cannot be reconstructed using this technique alone.}
All the above mentioned applications allowed us to construct a metric space for comparing synthetic SN~Ia spectra with observations.

\end{abstract}


\section{Introduction}
\label{sec:intro}

%
%

Type Ia Supernovae (SNe~Ia) are among the most luminous transients in the
Universe. They appear to be a rather homogeneous group, both
photometrically and spectroscopically. After the discovery of a
relation between their light-curve shape and luminosity
\citep{1993ApJ...413L.105P} {and of the luminosity-color relation \citep[e.g.][]{1996ApJ...473...88R,1998A&A...331..815T}} they have served as ``standardizable candles'' and
distance indicators in cosmology.  This increases the need to identify
their progenitors and to understand the explosion mechanism. It is widely
accepted that SNe~Ia are the result of the thermonuclear explosion of a
white dwarf in a binary system, where the companion star is needed to trigger
the explosion. However, the nature of the companion
star, whether it is another white dwarf \citep{1984ApJS...54..335I,1984ApJ...277..355W} or a non-degenerate companion \citep{1973ApJ...186.1007W,1982ApJ...253..798N} is still an open question.
In these two scenarios, models differ from each other by 
 the amount of mass gathered by the primary white dwarf at the time of 
explosion, the mode of thermonuclear combustion or the ignition mechanism \citep{2000ARA&A..38..191H,2012NewAR..56..122W,2013FrPhy...8..116H}.

SNe~Ia spectra, albeit quite homogeneous, exhibit a non-negligible diversity of spectral features {\citep[e.g.][]{2005ApJ...623.1011B,2006PASP..118..560B,2006MNRAS.370..299H,2012NewAR..56..122W}}.
Studying  their spectral differences is a promising way to shed some light  on questions regarding their nature. 
{There are many} ongoing observational campaigns like the Nearby Supernova Factory \citep[SNfactory,][]{2002SPIE.4836...61A}, the Palomar Transient Factory \citep[PTF,][]{2009PASP..121.1334R} or the Public ESO Spectroscopic Survey of Transient Objects\footnote{\url{http://www.pessto.org}} \citep[PESSTO, e.g.][]{2013MNRAS.431L.102M} and a large number of SN spectra collected by the CfA Supernova Data Archive \citep{2012AJ....143..126B}, {the CSP sample \citep{2013ApJ...773...53F}, the Berkeley sample \citep{2012MNRAS.425.1789S}, and SN catalogs as SUSPECT \footnote{\url{http://www.nhn.ou.edu/~suspect}} and WISEREP \citep{2012PASP..124..668Y}.} The number of
well-observed SNe~Ia has become large enough to allow for a quantitative statistical analysis of their spectral and photometrical diversity. Likewise, the complexity and diversity of synthetic spectra have increased \citep{2013FrPhy...8..116H}, for the first time producing enough synthetic data to allow a coherent comparison between theoretical predictions and observations, although such a deep investigation is still to be reported.

In order to explore this potential, we aim at developing an enhanced framework where all information stored in a particular data set can be automatically used to characterize a given synthetic spectrum. This new \textit{metric space} was constructed using an extended version of the Principal Component Analysis (PCA) method.  
PCA has been successfully used to classify QSO spectra \citep[e.g.][]{{1992ApJS...80..109B},{1992ApJ...398..476F},{2004AJ....128.2603Y},{2006ApJS..163..110S}}, and it has become a standard technique in that field. 
It  is also widely used  for studying galaxy spectra \citep[e.g.][]{{1995AJ....110.1071C}} and stellar spectra \citep[e.g.][]{{1983A&AS...51..443W},{1998MNRAS.298..361B}}.
A non-linear extension of PCA has also been used to photometrically classify SNe, in anticipation of the comparatively scarce spectroscopic resources to be faced by future cosmological surveys \citep{2013MNRAS.430..509I}. Standard linear PCA was applied to SN~Ia spectra recently
by \citet{2006MNRAS.370..933J} and 
 \citet[][]{2011MNRAS.410.2137C}.
Both papers concluded that PCA can be useful to study the diversity among SN spectra once larger samples become available.

In this work, we use an Expectation Maximization PCA (EMPCA) algorithm as implemented by \citet{2012PASP..124.1015B}, which is capable of handling missing data and measurement uncertainties. The potential of information extraction enclosed in EMPCA was enhanced by  pre-processing filtering and derivative routines, as well as by the use of complete spectral sequences in the construction of the initial data matrix. Once a stable PC space was obtained, we used Partial Least Square (PLS) analysis to demonstrate that the information it contains is not restricted to spectral indicators (velocities and pseudo-equivalent widths) but, as expected, it also correlates with photometric features as SALT2 \citep{2007A&A...466...11G} parameters $c$ and $x_1$. The outcomes from this analysis, applied to data from SNfactory, enabled the construction of a metric space where any given synthetic spectrum can be projected and automatically confronted with real data ones.
{Systematic comparisons of models with observation have been explored \citep[e.g.][comparing light-curves]{2013ApJ...773..119D}. Here we approach the problem from a new observation-driven perspective and we focus on spectral series.}

The paper is organized as follows: in Section  \ref{sec:method} we present details about all pre-processing techniques and statistical methods used to build our framework. The method is presented as a general data analysis technique, which allows its application to any set of spectral sequences. The connection with SN Ia data is presented in the following sections. Section \ref{sec:app} describes the SNfactory data set and the additional spectroscopic and photometric features to be investigated through PLS algorithm. Results from the EMPCA analysis \citep[][]{2012PASP..124.1015B}, including illustrative comparisons to models, are presented in Section \ref{sec:res_empca}. Section \ref{subsec:obs} presents the independently measured SNe~Ia features investigated in this work and the corresponding results from PLS are shown in Section \ref{sec:res_pls}. Finally, our conclusions are delineated in Section \ref{sec:conclusions}.

\section{Method}
\label{sec:method}

\subsection{Weighted Savitzky-Golay filter}
\label{subsec:savgol}

Before attempting any process of information extraction on spectral data, one must  take into account the high impact of random noise originated in the observational process. Spectra are affected by noise arising from photon statistics, detectors, and calibration.
Ideally, we would like to extract the features filtering the noise without degrading the signal.

The Savitzky-Golay (SG) filter \citep{1964AnaCh..36.1627S} is sometimes used to tackle this issue \citep{2012PASP..124.1015B, 2010ApJ...721..956P, 2007ASPC..372..249H}. It uses a least-square approach to  fit a polynomial to neighbouring points within a fixed window around each wavelength.  
 In comparison with other smoothing  methods (e.g. simply re-sampling the data in larger wavelength bins), the SG filter, with an appropriate choice of parameters, is more successful in preserving the shape of the peaks and valleys, even for weak spectral features.
The procedure is effective especially if the line broadening is significantly larger than
the size of the wavelength bin  as is the case here. Ideally, the smoothing window (polynomial degree) should be chosen such that it is not too small (large) to fail to filter the noise at the same time that it is not too large (small) so weak features are completely wiped away.

In this work, we wish not only to properly smooth a noisy spectrum, but we look for a procedure that takes into account the uncertainties associated with each measurement. Moreover, we should be able to calculate all the coefficients of the polynomial fit as well as their covariance matrix. 
In order to fulfil these requirements, we substituted the least square polynomial fit in the standard SG filter, by a weighted least square routine\footnote{\url{http://docs.scipy.org/doc/numpy/reference/generated/numpy.polyfit.html}}, where the quantity to be  minimized is given by
\begin{equation}
S=\sum_{i=1}^N\left[w_{i}\left(F_{\lambda_i}^{\rm obs} - g_{\rm M}(\lambda_i,\pmb{\beta})\right)\right]^2.
\end{equation}
Here, $N$ is the number of data points included in a fixed window, $F_{\lambda_i}^{\rm obs}$ is the observed flux at wavelength $\lambda_i$, $g_{\rm M}$ is the polynomial of degree M, $\pmb{\beta}$ is the vector of scalar coefficients of $g$ and $w_i$ is the weight assigned to $F_{\lambda_i}^{\rm obs}$ . The algorithm returns the best fit values and covariance matrix for $\pmb{\beta}$ at each wavelength. The width of the window is kept constant in $\log(\lambda)$, which corresponds to a constant velocity broadening to allow for a 
reasonable smoothing up to the minimum line broadening of the lines. 
{Although other types of smoothing techniques might possibly improve the results of our analysis, this matter  has not been investigated in detail in this work.}

Once the impact of noise is reduced, we proceed to the construction of a framework capable of extracting information from a large data set, while minimizing the number of random variables to be dealt with. 

 \subsection{Expectation Maximization PCA}
\label{subsec:empca}

Principal Component Analysis (PCA) is a dimensionality reduction method used to describe an initially multivariate data set using a smaller number of uncorrelated parameters (principal components --- PC). 
 It transforms the original high-dimensional space, through  a rotation of its axes. 
The first new axis (or PC) is aligned  with the direction
of largest variance in the data. The second PC should also maximize the variance, subject to being orthogonal to the first, and so on. Mathematically, these directions can be more easily determined through the  covariance matrix,  
\\
\begin{equation}
\Sigma_{ii'} = \frac{\sum_{k=1}^{k=N}{(X_i^k-\overline{X_i}) (X_{i'}^k-\overline{X_{i'}})}}{N},
\label{eq:cov_matrix}
\end{equation}
where $\overline{X_i}$ is the mean of all fluxes measured at wavelength $i$ and $N$ is the total number of objects {\citep[for a complete review, see][]{Jollife2002}}. Hereafter, we will always refer to the initial data as the mean subtracted terms in Eq. \ref{eq:cov_matrix} (the centralized version of all points in the initial data set).  

Once $\Sigma$ is diagonalized, the PCs are given by its eigenvectors, with the first PC corresponding to the one with the largest associated eigenvalue and so on. We are now able to fairly reconstruct a given spectrum from the original data set using only $M$  PCs ($M\ll N$), 
\begin{equation}
\pmb{F}_{\rm rec}\approx\overline{\pmb{X}}+\sum_{j=1}^{M}c_j \pmb{P}_j,
\label{eq:rec}
\end{equation}
with $\overline{\pmb{X}}$ representing the mean of all spectra, $\pmb{P}_j$ the $j$-th PC and $c_j$ the $j$-th scalar whose values must be determined from fitting $\pmb{F}_{\rm rec}$ to the measured flux. Geometrically, $c_j$ represents the projection of the measured spectrum on $\pmb{P}_j$.
{PCA is just a basis change. Using all the $N$ components the reconstruction becomes identical to the original data. The point is that the new basis captures a large fraction of the variance in a small number of components $(M)$. } 
 For the purpose of this work, the determination of the {``optimal''} $M$ is not a crucial point. A deeper discussion and other important applications of PCA for reconstruction in astronomy can be found in \citet{ishida2011, ishida2011b, benitez2012, 2013MNRAS.436..854B} and references therein.


If a particular measurement is missing, or is not reliable enough to be considered on the same basis as the other more accurate ones, it is possible to
reconstruct it from the nearest ones. Here we chose 
a different approach, taking advantage of a technique able to deal with 
missing elements in the initial data matrix: an expectation maximisation algorithm of PCA, first developed by \citet{roweis1998algorithms}. We use an extended version of it, which can deal with   non-uniform errors in the known components \citep{Dempster77maximumlikelihood,2012PASP..124.1015B}.

Reversing the line of thought which leads us to equation~\ref{eq:rec}, we can think of the PCs as the vectors which minimize $\chi^2=\sum_{k=1}^N\left[\pmb{X}^{k}-\pmb{F}_{\rm rec}\right]^2$.  In the presence of measurement errors, one can add a $k \times i$ weight matrix, $\mathbf{W}$, which controls the degree of influence of each flux measurement (for object $k$ at wavelength $i$) in the determination of the components,   
\begin{equation}
\chi^2=\sum_{k=1}^N\mathbf{W}^k\left[\pmb{X}^{k}-\pmb{F}_{\rm rec}\right]^2.
\label{eq:chi2}
\end{equation}
The above expression presents the challenge of diagonalizing a possibly very large matrix with a  non-negligible number of null elements. Within EMPCA, this problem is tackled through the use of an \textit{Expectation Maximization} algorithm 
{\citep[explained in detail in section 5.3 of][]{2012PASP..124.1015B}.}:

\begin{algorithm}
\caption{Expectation Maximization algorithm}
\begin{enumerate}
\item $\mathbf{V} \leftarrow$ random {orthonormal basis } of dimension { $i\times M$}
\item repeat until convergence {(i.e. the basis V does not}

 \ \ \ \ \ \ \ \ {vary significantly with new iterations)}:

 \ \ \ (a) calculate the projections of all spectra on {the basis}

 \ \ \ \ \ \ \ \ $\mathbf{V}$ (E-step)

 \ \ \ (b) using these coefficient values, find {a new estimate}

 \ \ \ \ \ \ \ \ {of the basis $V$} which
 minimizes equation \ref{eq:chi2} (M-step)

 \ \ \ (c) normalize the {columns of $V$} to unit length

\item return $\mathbf{V}$ {as the EMPCA calculation of the first}

 \ \ \ \ \ \ {$M$ eigenvectors of the basis $\mathbf{P}$} 

\end{enumerate}
\label{algo:EM}
\end{algorithm}

This method allows us to perform PCA on real data by giving higher weight to points with lower noise. Moreover, missing components in the input data are handled easily by assigning them  a weight equal to zero. 
Using the SG filter and EMPCA, we are able to  translate a set of spectra from wavelength to PC parameter space, with the 
SG filtering being crucial to ensure stability of the EMPCA results.
In the absence of such filtering, the EMPCA procedure does not converge to a stable solution.

\subsection{Error budget}
\label{subsec:error}

The propagation of the errors from the spectra to the projections  is not included in the EMPCA framework.
For standard PCA, the error in the determination of each eigenvector is inversely proportional  to the corresponding eigenvalue \citep{Jollife2002}. 
In EMPCA however, we need to deal with {three} main sources of error when analysing the geometrical distribution of our data in PC space. First, the iterative nature of the EM algorithm prevents us from obtaining the complete eigensystem and leads to uncertainty in the determination of the PC themselves. Beyond that, in the presence of missing data, computing the eigenvalues can be complicated, as it would require defining the total covariance based on an incomplete data sample. Second, once the PCs are given, we need to tackle properly the potential variance in their projections  due to missing elements in the data vectors. 
{Third, the variance in the projections due to noise.}

The determination of the PCs in EMPCA starts with  a
random first guess. It  rapidly converges to an approximate final
solution, but continues to  fluctuate weakly even after many more iterations.
The output PC vectors also vary slightly  for different  choices of the initial random seed. Despite the small influence of these features in the overall behaviour of our results, we took them into account by running the EMPCA algorithm  for 100 different seeds during 500 iterations each. The resulting sets of vectors were then used to estimate the uncertainty in the projections in PCs space. 
A small value of these variances can be interpreted as evidence that the input data quality is high enough to allow a stable determination of the PCs.

The errors in the  projections due to  missing measurements in the projected vector  were calculated assuming that the eigenvectors are well determined{, using the approach of \cite{Nelson20061}}. The propagation of the errors is due to the operation of projecting a non-complete spectrum on the PC space.
{The approach involves the inversion of submatrices of the covariance matrix,} whose dimension is much larger than the sample size. 
{An estimate of this matrix} was achieved by completing the observed data with the PCA reconstructions. Then, we computed the estimator for the covariance of the completed data as described by \citet{ledoit2004well}.  With the covariance matrix and the eigenvectors we computed the error in the projection due to missing data for each object, as described by \citet{nelson2002treatment}, section 3.2.1 and \cite{Nelson20061}. 

The errors on the projections due to measurement noise were computed using a Monte Carlo approach. 
Each spectrum was submitted to the SG filter and a random noise based on the original error amplitude was added to the smoothed spectrum. The new noisified spectrum was again submitted to the filtering process and its corresponding projection in PC space was computed. The procedure was repeated 25 times. This allowed us to assess, in an empirical approach, the variance in the projections due  to different magnitudes and covariances among the measurement errors. 

\subsection{Optimizing information extraction}
\label{subsec:opt}

After the smoothing described in Section \ref{subsec:savgol}, we are left with a well behaved representation of the measured spectra.  Mathematically, this would be enough to feed the EMPCA algorithm and perform the exercise of looking for patterns/subgroups in PCs space \citep[e.g.][]{{1983A&AS...51..443W},{1992ApJ...398..476F},{1995AJ....110.1071C}}. However, astronomical spectra commonly also present uncertainties in large wavelength modes due to reddening, calibration problems, and on the absolute flux itself due to poor estimates of the distance of nearby galaxies. They can also present uncertainties on small wavelength modes due to CCD fringing at higher wavelengths, discontinuities in the overlapping region between spectra obtained with different spectrographs, or poor subtraction of telluric lines. In this context, our goal is to optimize the power of information extraction as much as possible, getting rid of any recognizable additional noise  and enhancing intrinsic spectral features which we know to be relevant for individual object characterization. 

\subsubsection{Derivative Spectroscopy}

Although we are aware that it is not possible to completely remove the effect of extinction in measured spectra, we can make it easier to handle by, first, using the logarithm of the flux as our initial data. As an example, consider a general reddening  law:

\begin{equation}
 F_{\log} = \log_{10}{F_\lambda^{\text{obs}}}= \log_{10}{F_\lambda^{\text{intr}}}-0.4 \frac{A_\lambda}{A_V} R_V E_{B-V}{\rm \textbf{,}} 
\label{eq:log_flux}
\end{equation}
where $F_{\lambda}^{\rm obs}$, $F_{\lambda}^{\rm intr}$ and $A_\lambda$ are the observed flux, intrinsic flux and extinction at wavelength $\lambda$, respectively. $A_V$ represents the extinction in $V$-band and $R_V = A_V / E_{B-V}$, 
 and $A_\lambda/A_V$ is traditionally used to characterize the dust responsible for the extinction.
From this expression we realize that in terms of $F_{\log}$, reddening becomes a linear relation in the extinction parameter, $E_{B-V}$. Moreover, two objects following  the same extinction law but subjected to different amounts of reddening will differ only by a multiplicative constant. 

We would also like to take full advantage  of the PCA dimensionality reduction power by equally weighting the information contained in weak/strong spectral lines. The presence of strong lines naturally dominates the variance (and consequently all results from PCA) of any given spectra data set. They are crucial to the initial classification, but in a second order analysis they may obscure  important information contained in weak spectral features, which are more sensible to the conditions  of the material because usually they are not saturated. It is important to emphasize that PCA itself is an excellent framework to study a ``forest'' of weak lines since this kind of study demands the parallel analysis of many of them. 

We independently rediscovered a technique used in chemistry since \cite{morrey1968determining}, which consists of beginning the analysis from the derivative of each spectrum over the wavelength, which in our case translates to $\partial F_{\log}/\partial \lambda$, hereafter $dF_{\log}$.
This approach presents a few important improvements over the standard scenario for spectra analysis with PCA:
\begin{itemize}
\item Weak lines are emphasized. PCA on the derivative accounts for variance in the slope instead of variance in the flux, which also enhances the importance of the velocity of lines.
\item It does not depend on errors in distances or on small calibration errors of each spectrum, since a change in any of these adds a constant to $F_{\log}$ but leaves its derivative unchanged. 
\item It is  only mildly dependent on reddening and large but smooth calibration errors, since these 
add a function to $F_{\log}$ which is weakly dependent on wavelength (section \ref{sec:res_empca}).
\end{itemize}

\subsubsection{Complete spectral sequences}

The procedure described up to now can be applied to any data set composed of at least one spectrum per object. In a few cases however, mainly concerning transients,  a specific data set will contain a sequence of spectra for each of its objects, taken at different epochs. When this is the case,  we could, in principle, restrict ourselves to a single important epoch which would mean wasting a large part of the available information.
Such a time-focused analysis would have no means of recognizing distinct evolutionary tracks for two objects which happen to present similar features at the chosen epoch. Similarly, it would overestimate the distinction among  two sources sharing almost identical spectral  time evolution, if they are submitted to external effects which are mainly detected at the time of observation (such as noise, or bad atmospheric subtraction). 

Alternatively, one could compare results from the  analysis of spectra taken at different epochs and follow the different PC space configurations over time. Although this naively seems a good option, it poses some difficult technical problems. Comparing PCA results from two different matrices would require spectra for all sources taken at exactly the same epochs (or within the same epoch bin) in order to have enough statistics to justify a PCA in each one of them. As this is not the case for current data sets, we chose to analyse all available spectra in a single PC space by concatenating subsequent spectra in each line of the initial data matrix. In this context, if one particular object is missing one spectrum  the corresponding slots for those measurements are assigned a null weight, and the EMPCA algorithm still uses the available data in the determination of the complete PC space.

\subsection{The Partial Least Square analysis}
\label{subsec:PLS}

We now have a few techniques enabling us to translate the measured spectra from wavelength into PC parameter space. 
This new optimized space summarizes the information contained in the original data, grouping  objects similar to each other and providing a low-dimensional basis from which we can reconstruct the {main aspects of} observed spectra.
 However, given that the PC space represents the essential information contained in each spectral sequence, it should be possible to obtain additional information from the PCs. It is reasonable to assume the existence of  correlations between physical characteristics and a space that represents all spectral features, and in such case, we would be able to associate known physical characteristics to the parameters found with EMPCA. In this context, we could easily recognize a missing or unexpected element in synthetic spectra.
In this sub-section we show how the PLS analysis is suited for this task.

The Partial Least Squares analysis (PLS, also known as Projection to Latent Structures) is a technique used to find hidden relations between two groups of variables, originally developed by \citet[][]{ {wold1982soft},{wold1984collinearity}}. The underlying hypothesis behind PLS is that all observed data are generated by a small number of latent variables, not directly observed or measured. It searches for traces of these latent structures which may be present in different parameter spaces.

We can roughly think of PLS as a combined principal component search. Suppose we have two independent sets of variables, $\{\mathcal{X},\mathcal{Y}\}$, which result from measurements performed on the same objects. For example, $\mathcal{X}$ can be  a set of spectra and $\mathcal{Y}$ the set of independently measured photometric properties of the same objects. If we apply PCA to each one of these sets individually, we would obtain two distinct groups of PCs and their corresponding data projections, but the PCs of $\mathcal{X}$ would bare no information about the PCs,  or projections, of $\mathcal{Y}$, and vice versa. The goal of PLS is to determine directions within $\mathcal{X}$ and $\mathcal{Y}$ that maximize the covariance between their projected data.  Once the directions are known, from measurements of a new object in $\mathcal{X}$ we can estimate its projections  and predict the values for variables in $\mathcal{Y}$.

In this work, we look for relations between a 1-dimensional parameter space $\mathcal{Y}$ and the $M$-dimensional PC space coming from EMPCA. Mathematically, we are searching for the direction $\pmb{e}$ ($\sum_{i}{ e_i^2} = 1 $) that maximizes 
\begin{equation}
Cov( \pmb{e}X, Y ) = \frac{\sum_{k=1}^N{ ( Y^k-\overline{Y})\sum_{i}{  (X_i^k-\overline{X_i} ) e_i } }}{N},
\label{eq:PLS_cov}
\end{equation}
where N is the number of objects and $\overline{X_j}$ and $\overline{Y}$ are means:
\[\overline{X_j} = \frac{\sum_{k=1}^N{X_j^k} }{N}, \  \overline{Y} = \frac{\sum_{k=1}^N{Y^k} }{N} .\]
The corresponding correlation is the covariance weighted by the variances:
\[ Corr( e_i ) = \frac{Cov( e_i )}{\sigma( \sum_{i}{ X_i e_i }) \sigma( Y ) },  \]
where
 \[ [\sigma( \sum_{i}{ X_i e_i })]^2 = \frac{\sum_{k=1}^N{ (\sum_{i}{ ( X_i^k-\overline{X_i}) e_i })^2}}{N}, \]
 \[  [\sigma(Y)]^2 = \frac{ \sum_{k=1}^{N}{(Y^k - \overline{Y})}}{N}.   \]

PLS does not maximize the correlation, as the standard least square linear regression does, because that would assign the same weight to all directions in $\mathcal{X}$.  Instead, it maximizes the covariance, which gives more weight to directions in $\mathcal{X}$ with larger variance (first PCs) and avoids overfitting problems. In this work, we use the PLS algorithm as implemented by the \textit{scikit-learn} statistical suite  \citep{scikit-learn}.

In principle it is possible to apply PLS  before the  PCA dimensionality reduction, however, given the large dimension of the original spectral sequence data, that would barely simplify the traditional approach. 
Moreover the EMPCA method allows us to deal with missing components and diverse weights, and consequently  apply the method to many more spectra without discarding incomplete or significantly noisy data.

\section{Application}
\label{sec:app}
In this section we apply the previously explained framework to SN~Ia spectra from the SNfactory. 

\subsection{The Nearby Supernova Factory}
\label{subsec:SNf}

{The SNfactory is an experiment carried out using the University of Hawaii 2.2m telescope, mounted at Mauna Kea. Its goal is to obtain a sample of well observed SNe~Ia in order to improve the measurements of cosmological parameters \citep{2002SPIE.4836...61A, 2006NewAR..50..436C}.  Spectra are acquired through a two-channel Supernova Integral Field Spectrograph \citep[SNIFS,][]{2004SPIE.5249..146L}},  which simultaneously covers channels $B$ (3200-5200 \AA) and $R$ (5100-10000 \AA). Discovery is largely automated {using images from the JPL's Near Earth Asteroid Tracker (NEAT) and from the QUasar Equatorial Survey Team with quantitative and traceable selection of SN candidates \citep{2007ApJ...665.1246B}. This removes biases induced by the reliance on existing galaxy catalogs.} Precise calibration is carried out in order to ensure agreement {with high-redshift SNe} \citep{2013A&A...549A...8B}. The spectra are deredshifted with independently measured host galaxies redshifts \citep{2013ApJ...770..107C}. Telluric lines are properly removed and Milky Way extinction corrections are applied to
all spectra {\citep{1998ApJ...500..525S}}. Each supernova is followed from before $B-$band maximum up to $40-45$ days after peak, resulting in  10-15 
flux-calibrated low resolution spectra for each object. Most of the observed SNe are at the low-redshift end of the smooth Hubble flow  ($0.03 < z < 0.08$), which enables a small error in the determination of distance from peculiar velocities while still being well within the homologous expansion regime.

Consequently, SNfactory provides a considerably large and relatively homogeneous data set of SNe~Ia spectra (151 SNe and 2323 spectra at the time of this analysis), ideal for the study of second-order features as the one proposed here. Since all spectra are obtained with the same instrument, resolution and host subtraction routine, the data set is homogeneous enough to allow for intrinsic astrophysical features to produce non-negligible effects in PCA results. In what follows, we shall directly probe this argument by correlating the remaining variance in flux measurements with specific photometric and spectroscopic SN features (Section \ref{sec:res_pls}).

It is important to emphasize that  we chose the SNfactory as a first test of these tools because the outcome would certainly {be} less obvious if obtained from a less homogeneous sample.  However, due to the incorporation of the SG filtering and the use of $dF_{\log}$, the method is flexible enough to be applied to a much more diverse SNe~Ia data \citep[e.g.][]{2012AJ....143..126B, 2012MNRAS.425.1789S}.

\begin{figure}
\begin{center}
\includegraphics[width=1.\columnwidth]{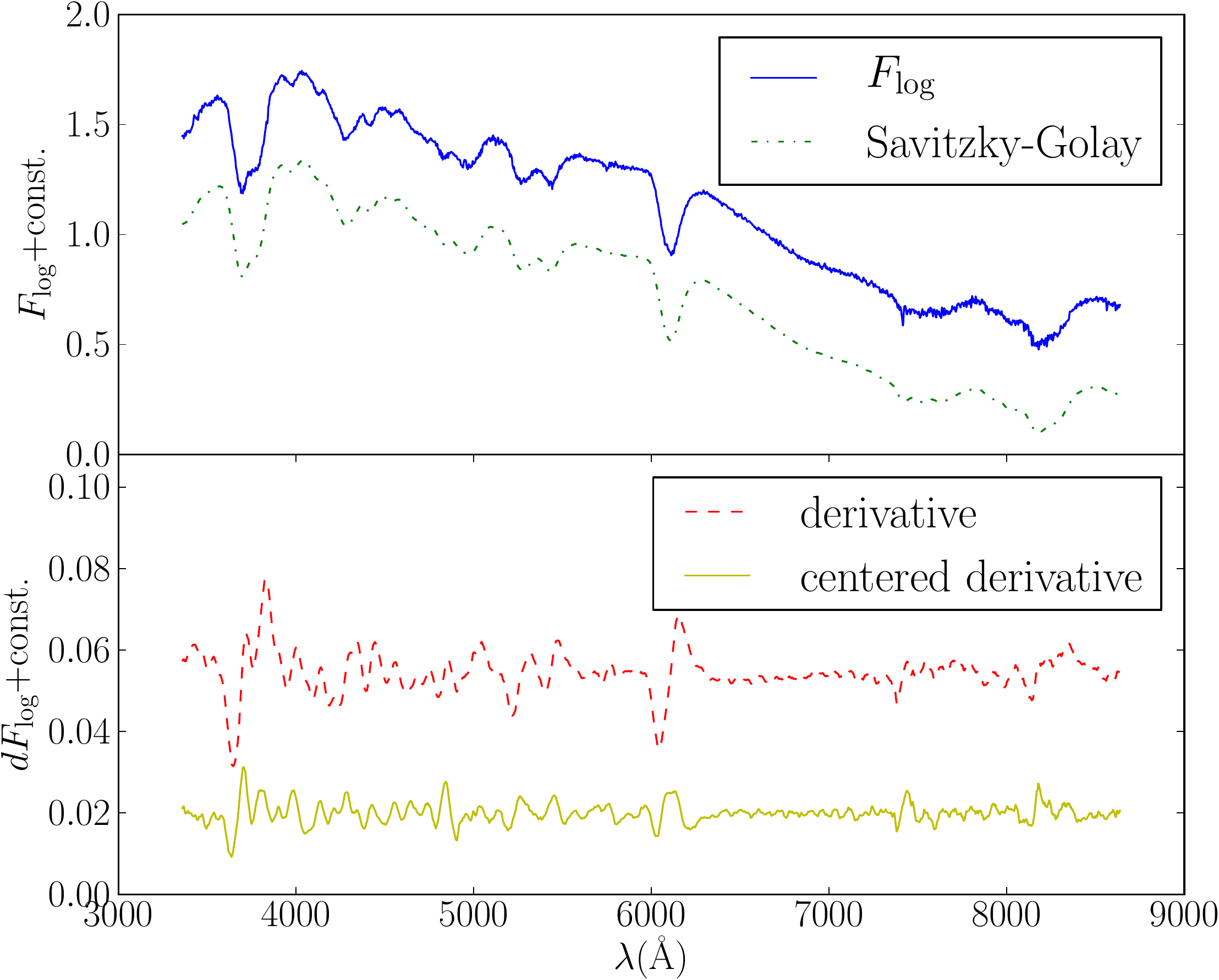}
\caption{Multiple steps in data treatment. Both panels show data from SNF20080626-002, taken at $-0.65$ days relative to  $B-$band maximum brightness. In each panel we artificially shifted the curves along the vertical axis for didactic reasons. \textbf{Top}:  $F_{\log}$ measurements before  (blue-full) and after (green-dotted) going through the SG filtering. \textbf{Bottom}: $dF_{\log}$ (red-dashed) and center derivative, $dF_{\log} - \overline{dF_{\log}}$ (yellow-full).}

\label{fig:data_steps}
\end{center}
\end{figure}

\subsection{Data treatment}
\label{subsec:datatreatment}

The processed portion of the  data set contains 151 SNe~Ia (2323 spectra) from which we selected objects with at least one spectrum before, one after $B-$band maximum and a minimum of three observed epochs between $-10$ and $+10$ days around $B$-band maximum. 
The epoch B-band maximum was determined from the SALT2 light curve fitter \citep{2007A&A...466...11G} applied on { magnitudes obtained from integrating $BVR$ top-hat filters \citep{2013A&A...554A..27P}}. Applying such requirements reduced our sample to {119 SNe and 764 spectra. The \Deltam\ of the sample is within $0.7$ and $1.7$, the SALT2 color within $-0.16$ and $0.40$. The redshifts of the SNe are within $0.007$ and $0.12$. Plots showing the distributions of these parameters in the SNfactory sample are shown by \cite{2011AA...529L...4C} and by \cite{2013ApJ...770..107C}}.

Each spectrum was smoothed by means of the weighted SG filter (Section \ref{subsec:savgol}), using  a third order polynomial (M=3), and a 6000 km/s-wide window as filter parameters. Those values were chosen by visually inspecting  some smoothed spectra; potentially, the choice of different values may provide further improvements.  
It is also important to emphasize that 
this filtering technique 
performs satisfactorily up to a certain threshold and starts to saturate for very noisy spectra. In this context, the uniformity and quality of SNfactory data allow us to apply the filtering without the need to discard spectra due to poor data quality.

\begin{figure}
\begin{center}
\includegraphics[width=1.\columnwidth]{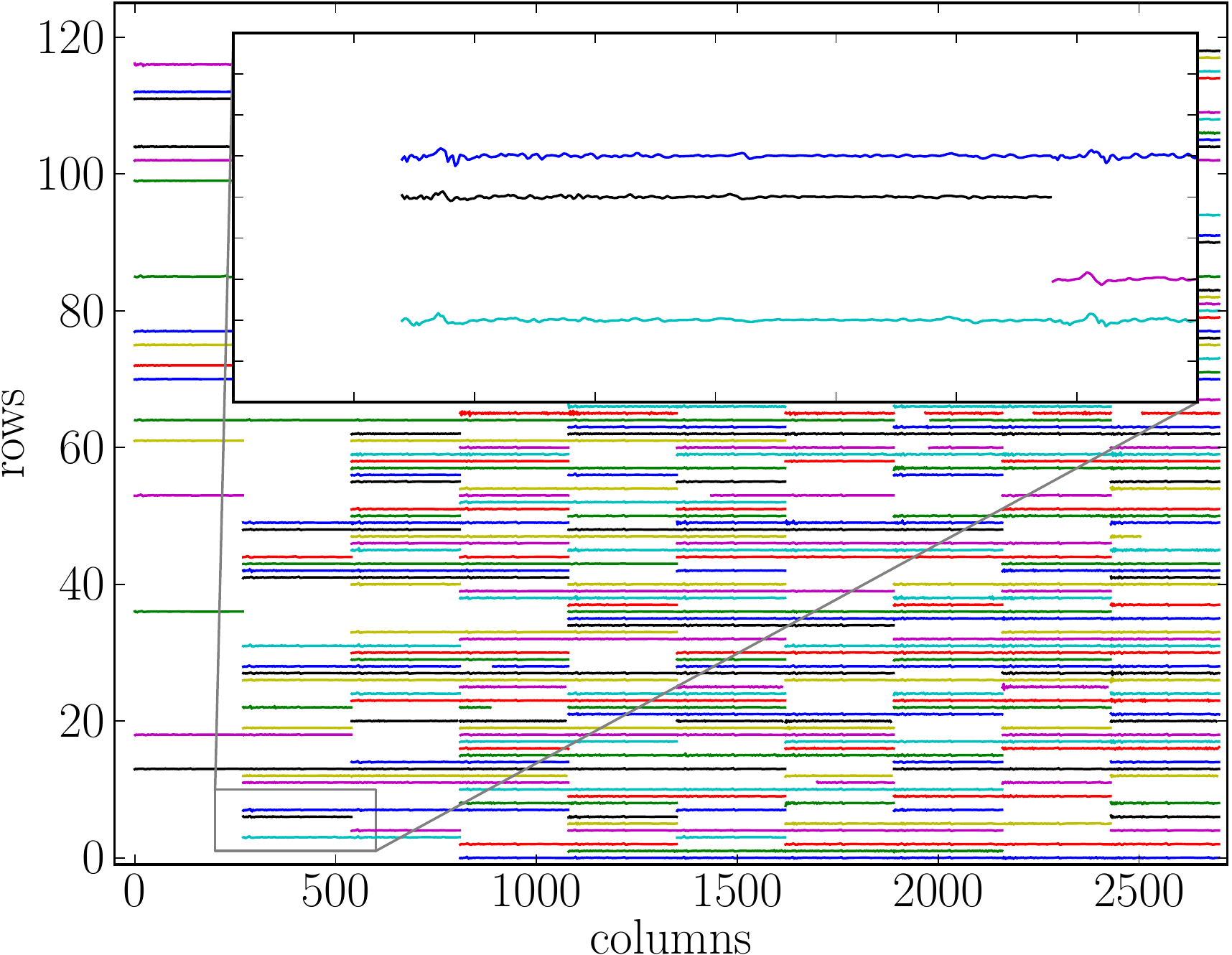}
\caption{Representation of the input data matrix. Different {rows} correspond  to  different SNe. Each column shows  centered  $dF_{\log}$, from spectra collected between $-10$ and $+10$ days relative to  $B-$band maximum brightness (from left to right), within each 2 day epoch window. Each curve runs over 3300\AA$\leq \lambda \leq$ 9000 \AA, written in wavelength bins of 20\AA. 
}
\label{fig:data_matrix}
\end{center}
\end{figure}

Figure \ref{fig:data_steps} is an example of how a measured spectrum is transformed at different stages of the pre-processing treatment. The top panel shows the measurements from the standard SNf reduction pipeline (blue-full) and the corresponding spectra after the SG filtering (green-dashed). The bottom panel presents the derivative of the same spectrum (red-dashed) and its centred counterpart (yellow-full), that is the difference between the derivative {and} the mean derivative. {The mean derivative is the mean of all SNe.} This last product of the spectra preparation was used as input to build the initial data matrix. In both panels, functions were artificially displaced along the vertical axis in order to improve clarity.

Once all preparations are done, each {row} in the data matrix is constructed by grouping into bins measurements taken in 2 days within each other. Thus, a SN with no missing spectrum is represented by a {row} in the data matrix constructed from the concatenation  of 10 spectra. The first was taken between $-10$ and $-8$ days, the second between $-8$ and $-6$ days, and so on. When a  spectrum is missing, its corresponding matrix elements are left empty, and if more than one measurement exists within the same epoch bin, the mean spectrum is used as a representation of that SN in that bin. 
{The choice of the parameters for the binning is inspired by the method of abundance tomography \citep{Stehle05,Mazzali08}. Using the SNfactory data, $-10$ days is as early as possible to have a rich sample. After $+10$ days the quality of the spectra generated by radiation transport codes not including forbidden line transitions starts to decrease \citep[e.g.][for a study of the SN~1991T]{2014MNRAS.445..711S}}.

As with the SG filter parameters, the size of the epoch bin can be adapted according to the characteristics of each data set. For SNfactory, a two-day binning is a reasonable compromise, given that SN~Ia spectra are quite homogeneous within this time frame and the data set is complete enough to provide a final matrix with more existing than missing spectra (in this configuration, we achieve 53\% coverage).  When transferring this procedure to another data set, one should  keep in mind that an epoch bin should be small enough to guarantee that spectral variations between different objects within that bin are not due to time evolution. At the same time, the bins must be large enough to accommodate uncertainties in the determination of the epoch for each spectrum and  allow a not too sparse initial data matrix.

Figure~\ref{fig:data_matrix} illustrates the overall shape of the final data matrix. Each spectrum was sampled every 20\AA\ (wavelength gap between two columns for the same spectra). Our results show that this choice has negligible effects on the analysis and saves computational time. 

In order to properly populate the weight matrix, errors coming from the flux measurements need to be propagated through  the filtering process.
 Since the complete error covariance matrix of each spectrum is not used in the EMPCA code from \cite{2012PASP..124.1015B},  we are computing only its diagonal terms.
The weighted polynomial fit described in Section \ref{subsec:savgol} represents the smoothed spectrum at each wavelength as
\begin{equation}
F_{\lambda}^{\rm obs} \rightarrow g_3(\lambda,\overline{\beta})=\beta_0 + \beta_1(\lambda - \lambda_0) + \beta_2(\lambda-\lambda_0)^2 + \beta_3(\lambda-\lambda_0)^3,
\end{equation}
where $\lambda_0$ is the central wavelength for each window. Given that each polynomial fit is used to determine the smoothed flux only at $\lambda=\lambda_0$, this implies that for each wavelength:
{\begin{eqnarray}
F_{\log}&=&\log_{10}F_{\lambda}^{\rm obs}\bigg|_{\lambda=\lambda_0}= \log_{10}{\beta_0},\\
dF_{\log}&=&\frac{d \log_{10}F_{\lambda}^{\rm obs}}{d \lambda}\bigg|_{\lambda=\lambda_0} = \frac{\beta_1}{\beta_0 \ln 10},
\end{eqnarray}}
{finally, propagating the errors:}
\begin{eqnarray}
\delta F_{\log}&=&\frac{\delta \beta_0}{\beta_0\ln 10},\\
\delta dF_{\log}&=&\left\vert\frac{\beta_1}{\beta_0 \ln 10}\right\vert\sqrt{\frac{\delta \beta_0^2}{\beta_0^2}+\frac{\delta \beta_1^2}{\beta_1^2}-\frac{2{\rm cov}(\beta_0,\beta_1)}{\beta_0 \beta_1}},
\end{eqnarray}
where $\delta \beta_i$ denotes the uncertainty associated with the determination of parameter $\beta_i$ and the covariance between the first two parameters is represented by ${\rm cov}(\beta_0,\beta_1)$ . The weight matrix elements are then defined as $w_i=\delta F_{\log}^{-2}$ or $w_i=\delta dF_{\log}^{-2}$ for the logarithm and derivative cases, respectively.

There are a few supernovae within the SNfactory set whose errors are an order of magnitude
smaller than the ones of the bulk of the data. 
This happens for bright SNe, where the number of counts is high and the Poisson error small. 
For example SN~2007le, being one of the nearest supernovae in the
sample, has errors much smaller than most of the other objects. If the EMPCA is carried out with errors as they come out of the SG filter, it would 
overweight the two or three supernovae with the smallest errors and the
first components would point in the direction of these few objects.  
This behaviour of EMPCA in the presence of few objects with a noise much lower than the rest of the sample is also highlighted by \citet[][section~8.3]{2012PASP..124.1015B}. 
To overcome this problem, we
artificially decreased the weight of 52 SNe (42\% of the sample) in order to have no SN with a weight larger than 90 times the sum of the weights of the other objects. Results  are not biased towards these objects and the PC space is stable as long as
their number is kept between $\sim25\%$ 
and $\sim75\%$ of the total data set. We also performed the analysis without changing the initial weights, but removing the 8 SNe with lowest noise from the initial sample. The test returned the same results, demonstrating the low sensitivity of this procedure regarding the method used for down-weighting. Once the PC space is determined, the spectra are not downweighted to obtain the projections.

\section{Principal Components Interpretation and Metric Space Comparison}
\label{sec:res_empca}

\begin{figure*}
 \begin{minipage}{.47\textwidth}
  \begin{center}
   \includegraphics[width=1.\columnwidth]{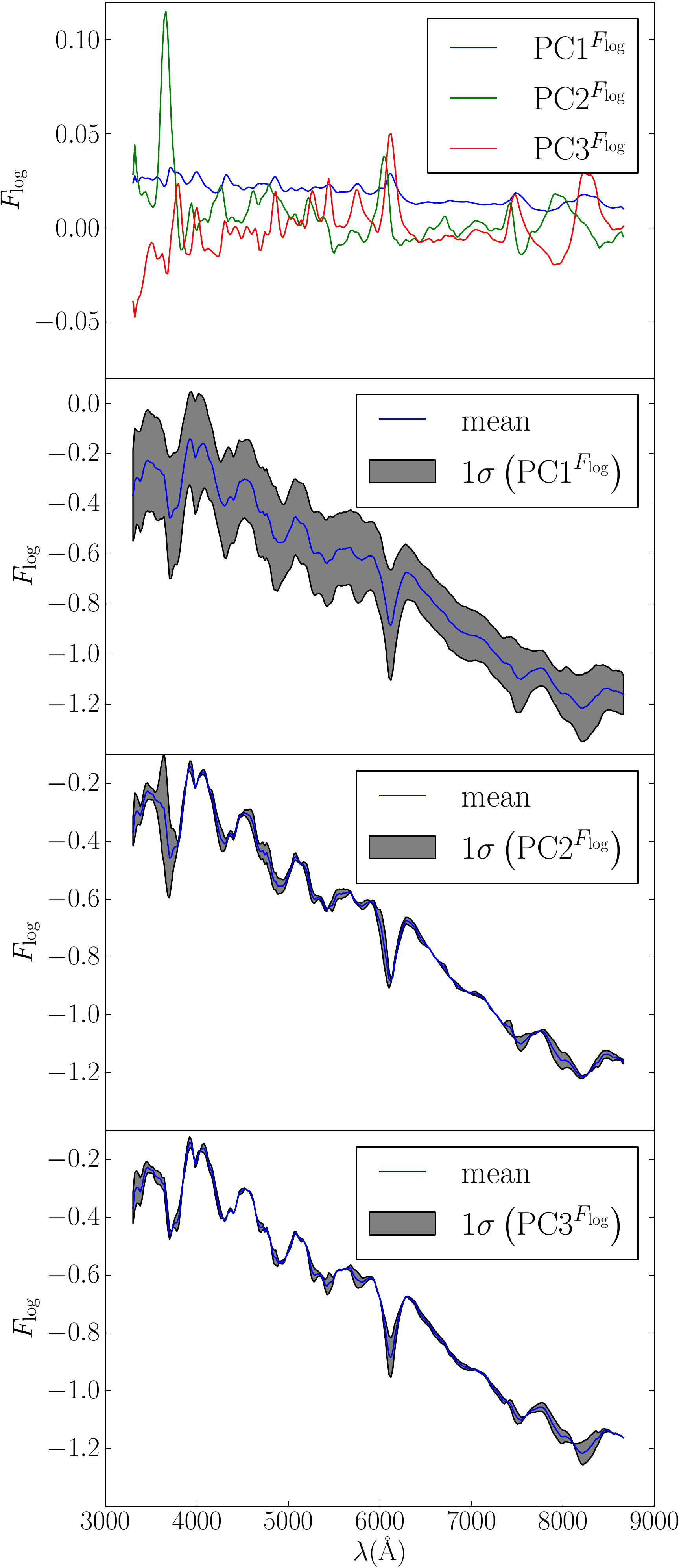}
   \caption{
   First panel shows the first three eigenvectors obtained from the analysis on $F_{\log}$.
   The second to fourth panels illustrate the main spectral features tracked by PC1, PC2 and PC3. All panels correspond to a  spectrum taken between $-6$ and $-4$ days relative to B-band maximum. Blue lines denote the mean spectrum. Gray regions were obtained by reconstructing the spectrum with only 1 PC and varying the scalar coefficient within the $1\sigma$ {range given by the data.      
   The PC2 and PC3 bare similarities with the Si and Ca components found by Chotard (2011)}.
    }
   \label{fig:vec_flux}
   \vspace{.025\linewidth}
  \end{center}
 \end{minipage}
 \hspace{.04\linewidth}
 \begin{minipage}{.47\textwidth}
  \begin{center}
   \includegraphics[width=1.\columnwidth]{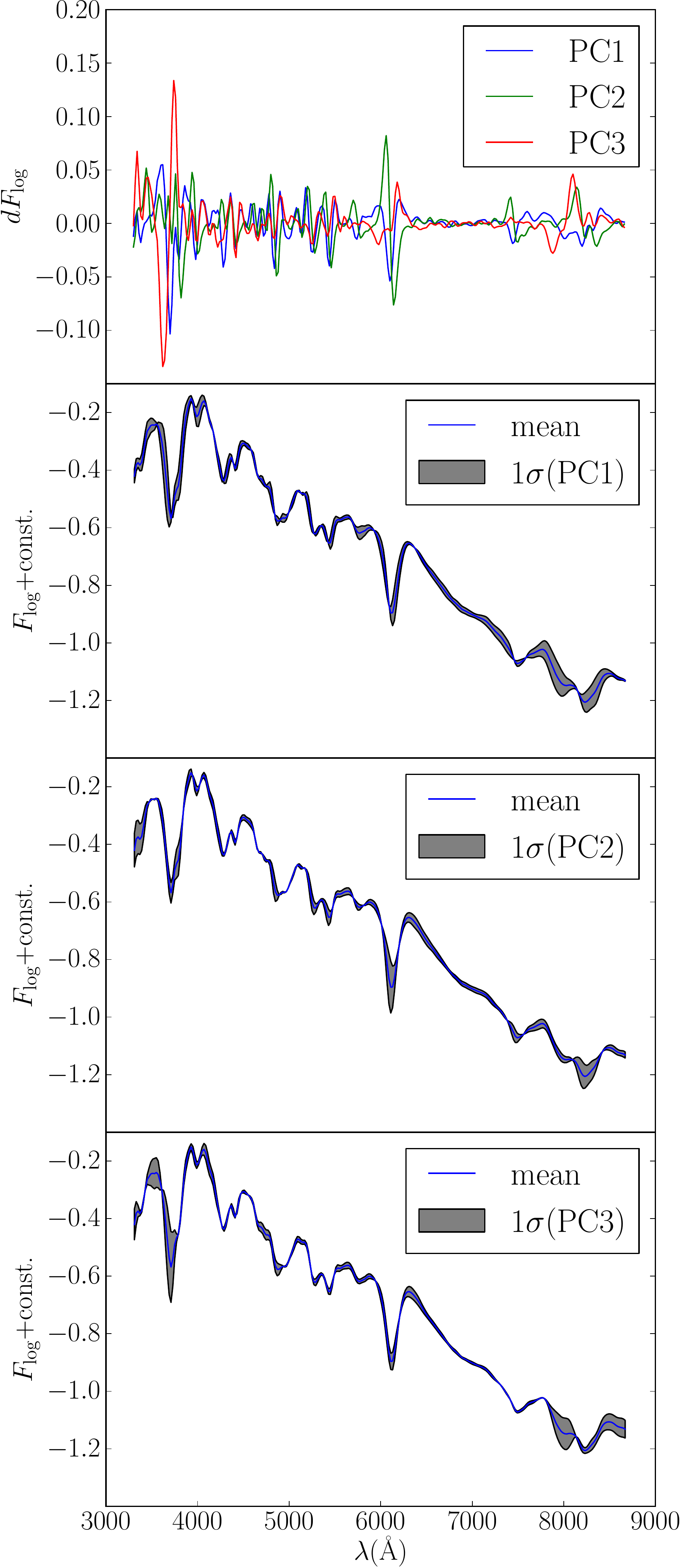}
   \caption{Same as Figure \ref{fig:vec_flux}, but from a data matrix based on $dF_{\log}$. \vspace{2.4cm}  } 
   \label{fig:vec_derivatives}
  \end{center}
 \end{minipage}
\end{figure*}

We present below, side by side, results from the application of the EMPCA to SNfactory data, with matrices built from $F_{\log}$ and $dF_{\log}$  (Figures~\ref{fig:vec_flux}~and~\ref{fig:vec_derivatives}, and Figures~\ref{fig:PCs_flux}~and~\ref{fig:PCs_derivative}). Hereafter, the PCs derived from a data matrix based on $F_{\log}$ will be referred to as PC$i^{F_{\log}}$, with $i$ denoting the PC number. Meanwhile, PC calculated from a matrix based on derivatives will be simply called PC$i$. This direct comparison allows the reader to clearly recognize the differences and advantages  in using the derivatives, which is a crucial step for the subsequent PLS analysis presented in Section \ref{sec:res_pls}.

\subsection{Principal Components}

Figures \ref{fig:vec_flux} and \ref{fig:vec_derivatives} show the behaviour (first panel) and consequent influence on reconstructed spectra (second to fourth panels) of the first three eigenvectors for analyses based on $F_{\log}$ and $dF_{\log}$, respectively. In both figures, the first panel displays the functional form of the PCs themselves, while the remaining panels show the effect we can achieve, in the final reconstruction, by increasing the weight assigned to each PC within the boundaries allowed by the data. The reconstructions presented here are non-cumulative. In other words, the gray region in each panel represents features which arise when combining the mean spectrum with each PC separately.
From this, we see that the first eigenvector computed from $F_{\log}$ (Figure~\ref{fig:vec_flux}) leads to a slow variation with wavelength in the reconstructed result. 
Its influence 
can be easily associated with a constant that allows  a rigid translation in flux, although is also carries some discrete wavelength dependent features. 
 Also, it clearly describes a much larger variance than the next two components (larger area covered by the gray region, second panel of Figure~\ref{fig:vec_flux}).
The first PC is largely influenced by dust, with its long wavelength behaviour 
being consistent with a Cardelli reddening law.
However, significant contributions to the flux and to the slope of this eigenvector due to absolute magnitude and intrinsic color variations are likely. 
The mixing of intrinsic and extrinsic properties is avoided by the PCA based on the derivative.
For $dF_{\log}$ (Figure~\ref{fig:vec_derivatives}, panels 2--4), one can notice that an important role is assigned to small scale variations. Moreover, the variance covered by the first PC  is comparable to that of the others. This is a direct consequence of our choice of removing the overall flux information from the input data through the use of the derivative. In this analysis, the first three PCs show variations of pseudo-Equivalent Widths (pEW) and velocities of many lines, some of which are studied in more detail in Section~\ref{sec:res_pls}.

\

\subsection{High velocity features}
\label{subsec:HVFs}
{The first two PCs contain a large part of the spectral variance in the SNfactory data. This will be studied in detail in the subsequent sections.}
We highlight the significant role played by the third PC shown in Figure~\ref{fig:vec_derivatives}, which tracks the variation of the High Velocity Features (HVFs) of \CaII~H\&K and infrared lines without particularly affecting the rest of the spectrum. {This figure represents the eigenvectors in the epoch range between $-6$ and $-4$ days relative to B-maximum, since the high-velocity part of these lines usually disappears at later epochs.}. 
 The third PC, by construction uncorrelated with the first two, seems to be mainly responsible for tracking variations of  HVFs of Ca. Thus, confirming that HVFs  of Ca is a property of the outer layers of the ejecta
and it is not correlated with the underlying structure \citep{2005ApJ...623L..37M}. Such an effect can be achieved with an asymmetric/clumpy outer layer of the ejecta convolved with  line-of-sight effects \citep{2006ApJ...645..470T} and is a good indicator of the kind of astrophysical characteristics which can possibly be recognized also in synthetic spectra.

\begin{figure*}
\begin{center}
\includegraphics[width=2.\columnwidth]{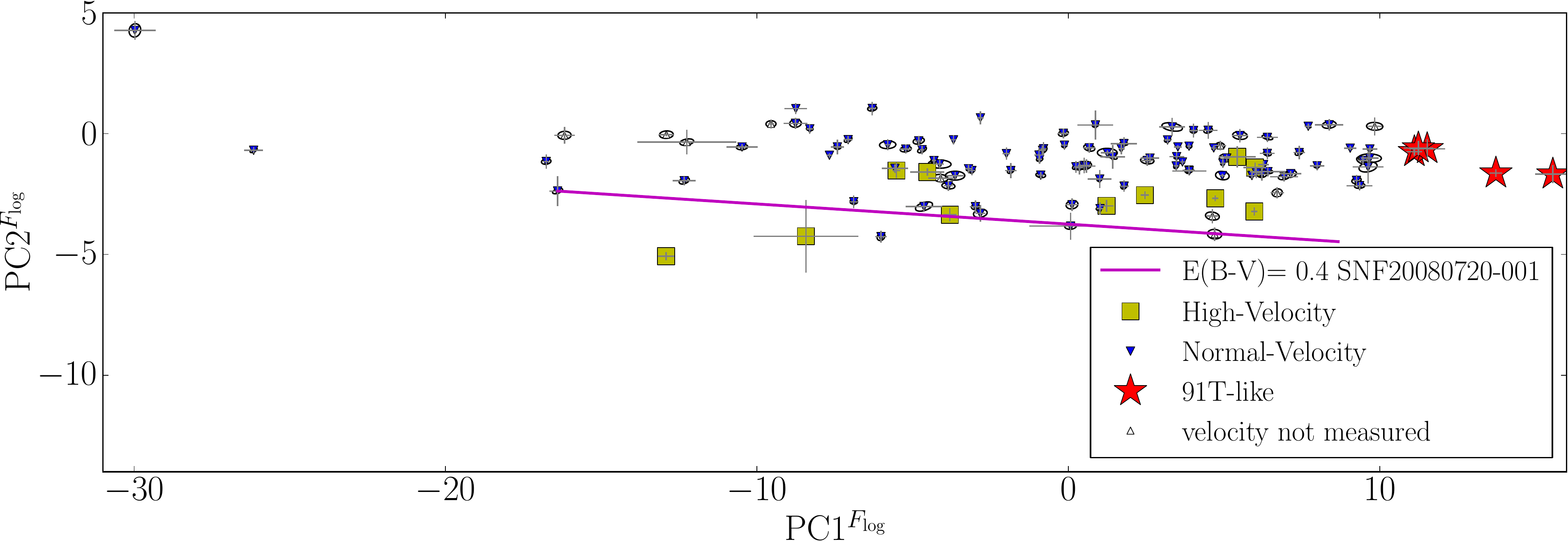}
\caption{Projections of SNfactory data on the first two PCs for analysis based on $F_{\log}$. Each point represents a supernova, colored according to the spectral classification of \citet{2009ApJ...699L.139W}. A few 91T-like SNe are also highlighted.
  The crosses correspond to $1\sigma$  errors coming from random seed variation and the ellipses denote $1\sigma$ uncertainties due to missing data  and measurement noise. The magenta line shows the effect of reddening on the projection of the SN SNF20080720-001, which presents an observed $B-V$ color of $\sim0.4$ mag.}
\label{fig:PCs_flux}

\end{center}
\end{figure*}

\begin{figure}
\begin{center}
\includegraphics[width=1.\columnwidth]{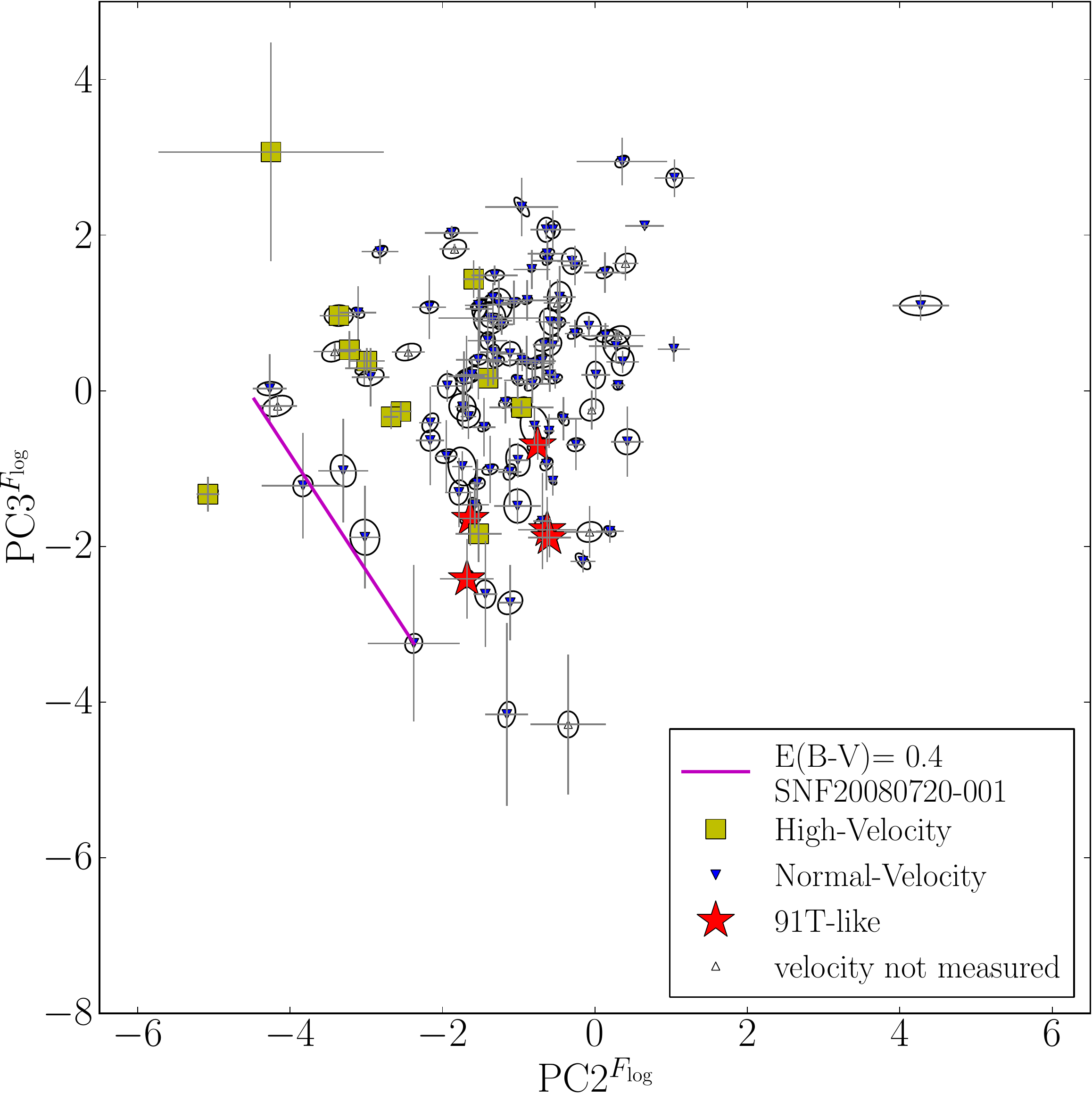}
\caption{Same as Figure \ref{fig:PCs_flux}, PC2$^{F_{\log}}$ and PC3$^{F_{\log}}$. 
}
\label{fig:PCs_2_3_flux}

\end{center}
\end{figure}

\subsection{Metric Spaces}

The projection of SNfactory data in a 2-dimensional PC space, obtained from $F_{\log}$, is displayed in Figure~\ref{fig:PCs_flux}. Individual objects are coloured  following the classification scheme  defined by  \cite{2009ApJ...699L.139W}, where high-velocity SNe are those whose velocity of the \SiII~6355~\AA\ is more than $3\sigma$ above its mean value. In what follows, we consider the mean~$+3\sigma$ equal to 12200~\kms, as computed by \cite{2012AJ....143..126B}. We also highlighted a few 91T-like SNe (red stars), {following the classification used by \cite{2012ApJ...757...12S}.  1999aa-like SNe are not highlighted as 91T-like.}
Crosses correspond to $1\sigma$ uncertainties due to random seed variation and ellipses represent the $1\sigma$ errors coming from missing data in the projected spectral sequence and measurement noise  added in quadrature. After exploring a large range of the MC parameters, {our results show that 25 realizations were more than enough to the secure stability of the error bars.} 

Figure \ref{fig:PCs_flux} can be considered to be an alternative visualization of the same effect as presented in Figure~\ref{fig:vec_flux}: the first PC obviously contains a larger part of the total variance, and consequently the interpretation of the subsequent PCs is obscured.  In this context, although we can identify a certain clustering of  91T-like SN in larger values of PC1, contamination is still significant, and an  attempt to separate the set according to these features would certainly present important drawbacks. 
This high level of contamination is  mainly due to reddening. This is shown clearly by the variation of the projections of SNF20080720-001 after a reddening correction of up to E$(B-V)=0.4$ with a \cite{1989ApJ...345..245C} law (magenta line in Figure~\ref{fig:PCs_flux}). This object has an observed $B-V$ color of $\sim0.4$, one of the reddest SNe in the SNfactory sample. Figure~\ref{fig:PCs_2_3_flux} shows the analogous situation for PC2~$\times$~PC3 parameter space. The magenta line corresponds to the reddening effect still present in the second and third PC in flux space, showing that the PCA in fluxes is not able to isolate the effect of reddening in the first PC.

\begin{figure*}
\begin{center}
\includegraphics[width=2.\columnwidth]{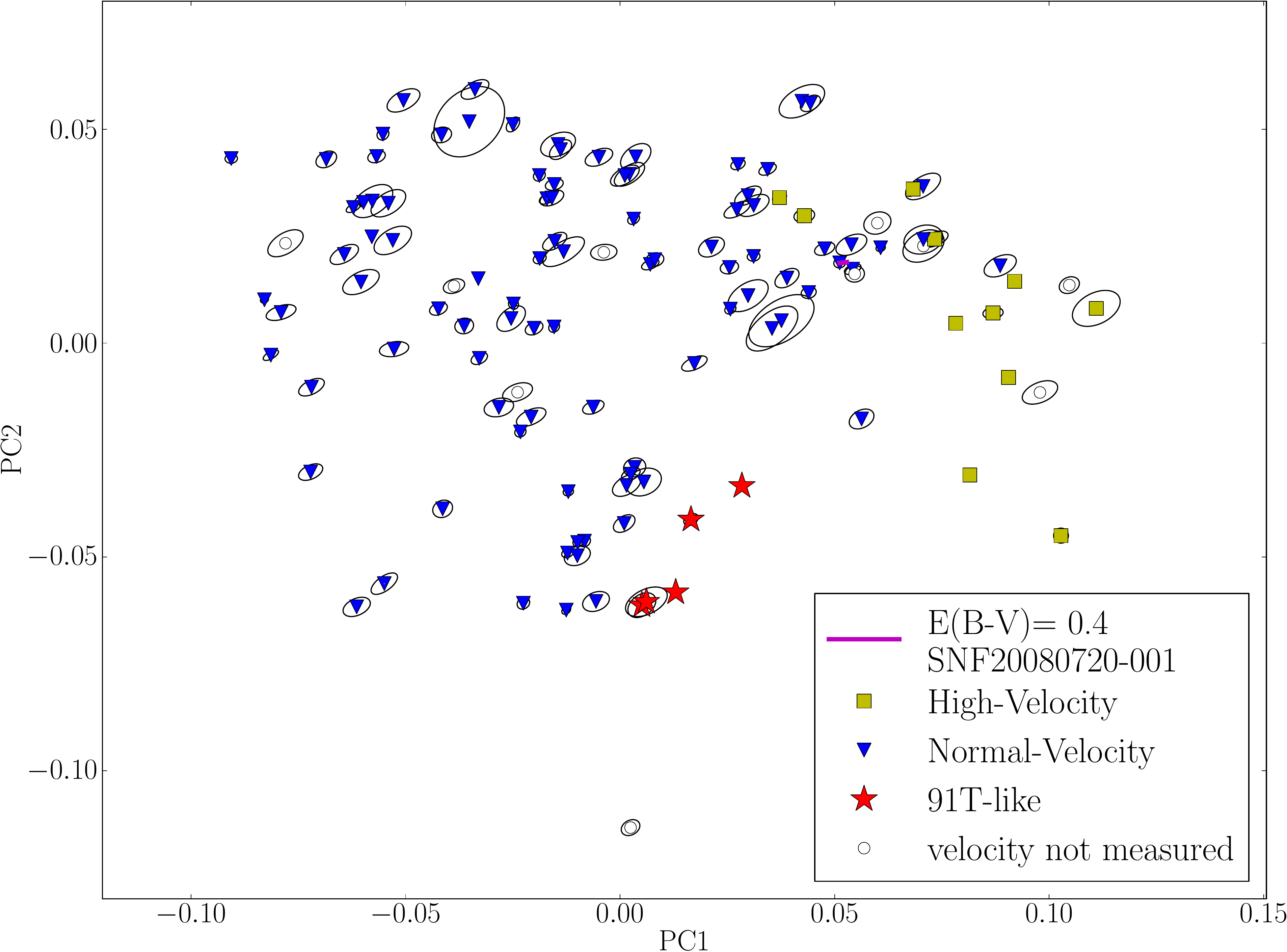}
\caption{Same as Figure \ref{fig:PCs_flux}, for the analysis based on $dF_{\log}$. 
}
\label{fig:PCs_derivative}
\end{center}
\end{figure*}

Figure \ref{fig:PCs_derivative} shows how this situation changes when the analysis is based on $dF_{\log}$. 
The crosses due to the instability of the EMPCA algorithm are completely negligible, the ellipses due to noise and missing components are large only in a few very noisy SNe.
  The slowly declining 91T-like SNe (red stars) are at the bottom edge of the diagram, clearly separated from the high-velocity ones on the right (yellow squares). The spectroscopically normal SNe (blue triangles) are spread throughout the parameter space, indicating a larger intrinsic variability between these objects.
 Visually inspecting  spectra from the SNe in the upper-left corner, we also realize that this space is occupied by fast declining SNe with cooler spectra showing a lower ionization ratio.
According to the projections in our metric space there are no clear separations that justify the definition of subclasses. SNe~Ia, accordingly to spectral features, look like a continuous distribution of objects.
In other words, there is no clear separation in velocity or EW of lines which justifies or objectively indicates a threshold for defining a subclass, 
although there are undoubtedly fundamental differences between objects in the extremes.
For example, 91T-like SNe show a ``bridge'' of objects that connects them with the bulk of normal ones. The same is true for the ones with a high velocity of Si.
 
The marginal effect coming from  reddening in this context is illustrated by the magenta line in Figure~\ref{fig:PCs_derivative}. As in Figure~\ref{fig:PCs_flux}, it represents the translation in PC space experienced by SN SNF20080720-001 when a 0.4 mag 
 reddening correction is applied.
Comparing the magenta lines in both figures demonstrates the power of the derivative analysis in minimizing the effect of dust in the PC space. Although this is one of the most reddened SNe, the change in the PCs is merely marginal. The same trend is observed for all the other objects in this sample.
 
It is important to keep in mind that
this specific geometrical  configuration in PC space will always be related to the sample of objects used to
construct it, and it is not a ``universal'' space for SNe~Ia.
However, it is reasonable to expect  that the addition of more high-quality data leads to an asymptotic PC space configuration which summarizes the similarities and differences within the SNe~Ia sample used in its construction.  
Nevertheless, with the SNfactory data at hand, we are already able to demonstrate  that the analysis is useful to look for correlations in
the data, attack the problem of SN~Ia spectra characterization and
search for outliers.

Although this ``universal" PC space is merely an asymptotic state, we can have a hint on how close it is to the ideal configuration. In other words, we can test the stability of a given PC space through the successive application of the EMPCA algorithm to different subsets of the original data. This procedure is called Cross-Validation (CV) and it has been used in many fields where the configuration of a given method depends on the initial data set \citep{2009arXiv0907.4728A}. Detailed results from a CV test are presented in Appendix \ref{ap:kfolding}, and these demonstrate the stability of the space presented in Figure~\ref{fig:PCs_derivative}.

After analysing the first pair of PCs and confirming the stability of the PC space, we are left with an obvious question: how many PCs are necessary to 
describe the data set and 
throw  away a substantial part of the noise?  In a standard PCA the fraction of the total variance associated to each PC, or to a subset of them,  can be estimated through the cumulative percentage of total variance \citep{Jollife2002, ishida2011, 2013MNRAS.436..854B}. Given that the eigenvalues associated with each eigenvector constitute a measurement of the data variance along that PC direction, this means that the ratio between the largest eigenvalue and the sum of all eigenvalues gives an estimative of the percentage of variance (or information) described by the first PC. However, in the EMPCA approach we do not have access to all eigenvalues at once, since the eigenvectors are calculated one at a time through the EM algorithm. Nevertheless, we do expect that only a handful of PCs will actually carry meaningful information and this hypothesis can be tested with a small sub-sample of them. 

We used the EMPCA approach to calculate the first six PCs and their corresponding data set projections. From these, we  determined the variance along each PC. By definition, the first PC contains a larger fraction of the total variance than any other PC, so we used it as a normalization factor. In this context, we can obtain an estimate of how much information is stored in a certain PC, in comparison to that in the first one. 

\begin{figure*}
 \begin{minipage}{.47\textwidth}
  \begin{center}
\includegraphics[width=1.\columnwidth]{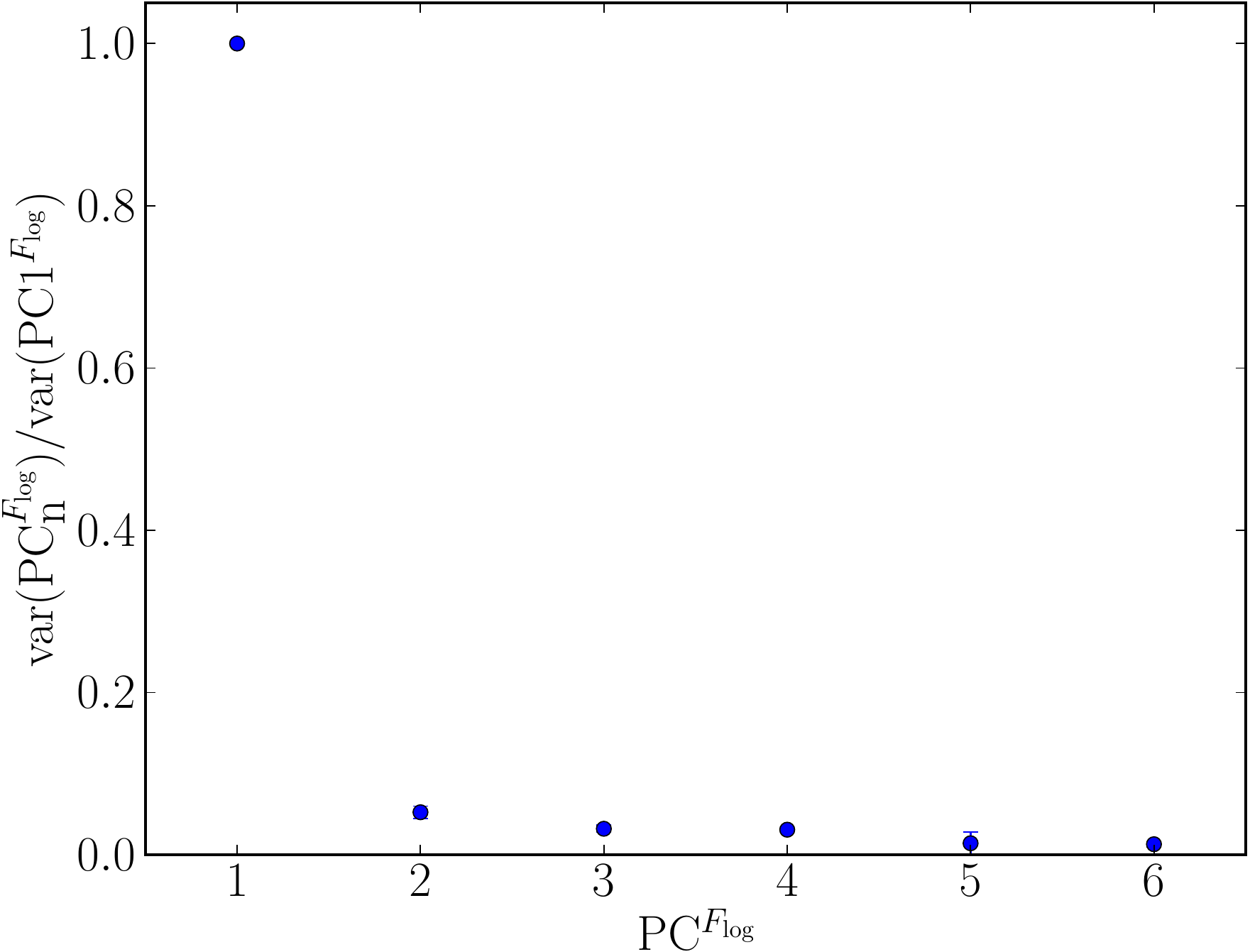}
\caption{Distribution of variance among the first 6 PCs from  ${F_{\log}}$ data matrix.
The variances are normalized to that of the first PC. {The errorbars show the variability due to k-folding (Appendix \ref{ap:kfolding}).}
}
\label{fig:variances_flux}
\end{center}
 \end{minipage}
 \hspace{.05\linewidth}
 \begin{minipage}{.47\textwidth}
\begin{center}
\includegraphics[width=1.\columnwidth]{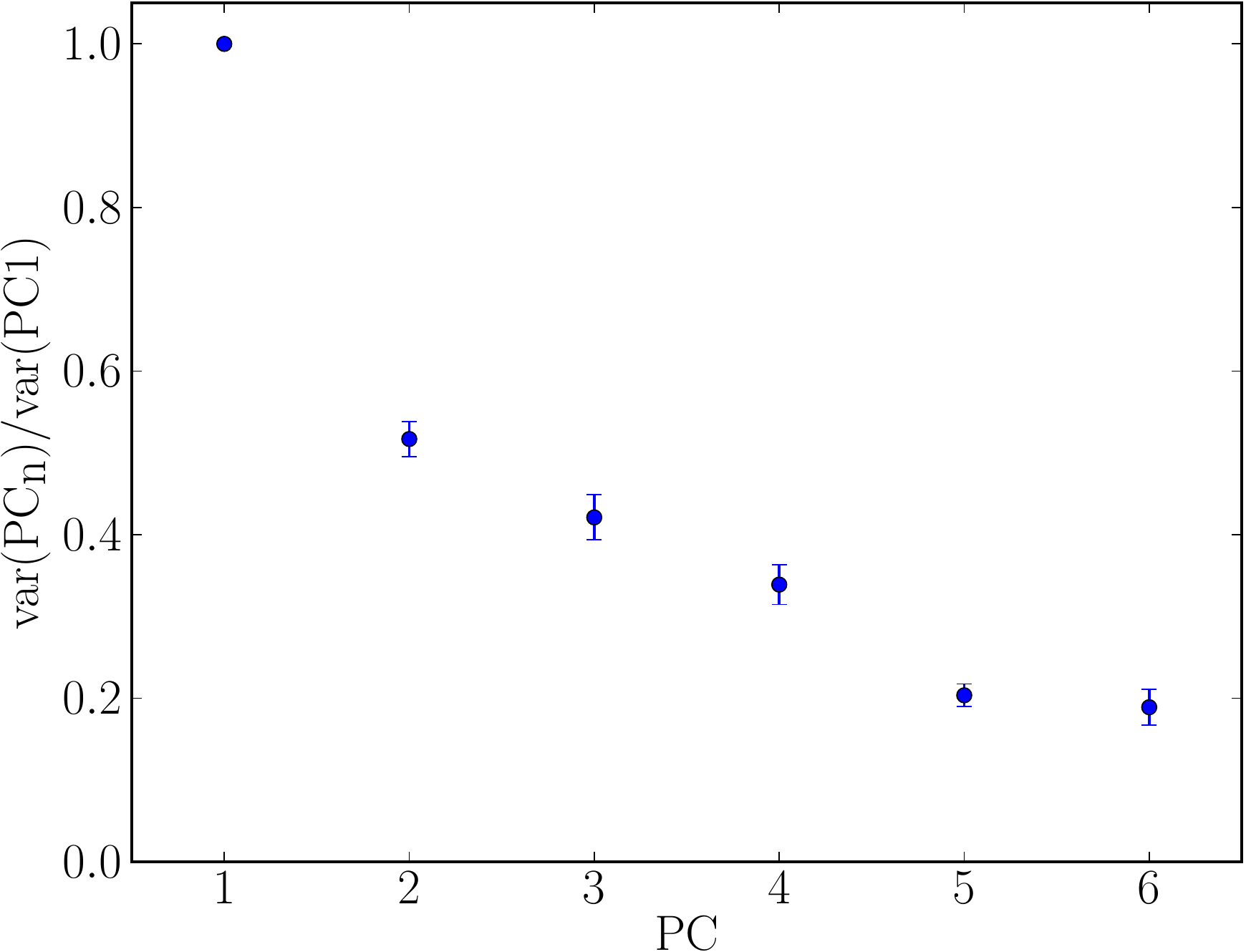}
\caption{Same as Figure \ref{fig:variances_flux}, but obtained from $dF_{\log}$ data matrix.\vspace{0.75cm}}
\label{fig:variances}
  \end{center}
 \end{minipage}
\end{figure*}

In Figures \ref{fig:variances_flux} and \ref{fig:variances} we show the variances normalized to the first component for the analysis on $F_{\log}$ and $dF_{\log}$ respectively.
Figure~\ref{fig:variances_flux} shows the same result we have seen in Figures~\ref{fig:vec_flux} and  \ref{fig:PCs_flux}, with most of the information concentrated in PC1$^{F_{\log}}$.
 From a physical perspective, performing the analysis in this parameter space is challenging,  due to extinction and intrinsic luminosity variations.
 Extinction effects are present in all the principal components, making it difficult  to disentangle two very different physical processes. 
 For example, in this context, two similar SNe subjected to different amounts of reddening would be distant from each other in the PC parameter space (as illustrated by the magenta line in Figures~\ref{fig:PCs_flux} and \ref{fig:PCs_2_3_flux}).
On the other hand, when using $dF_{\log}$ we concentrate the investigation on spectral features which are crucial to SNe~Ia characterization  
 and consequently a larger number of PCs are found to be significant.
The derivative approach removes the effect of reddening, a physical process that causes a large amount of variance in the data, making it easier to train the PCA space.
 From Figure~\ref{fig:variances},  it is clear that PC2 to PC5 carries at least 20\% of the variance in PC1 each and the fractions stabilized
 for PC6. Thus, we conclude that 5 PCs are enough to describe most of the  variance in SNfactory.

\begin{figure*}
\begin{center}
\includegraphics[width=2.\columnwidth]{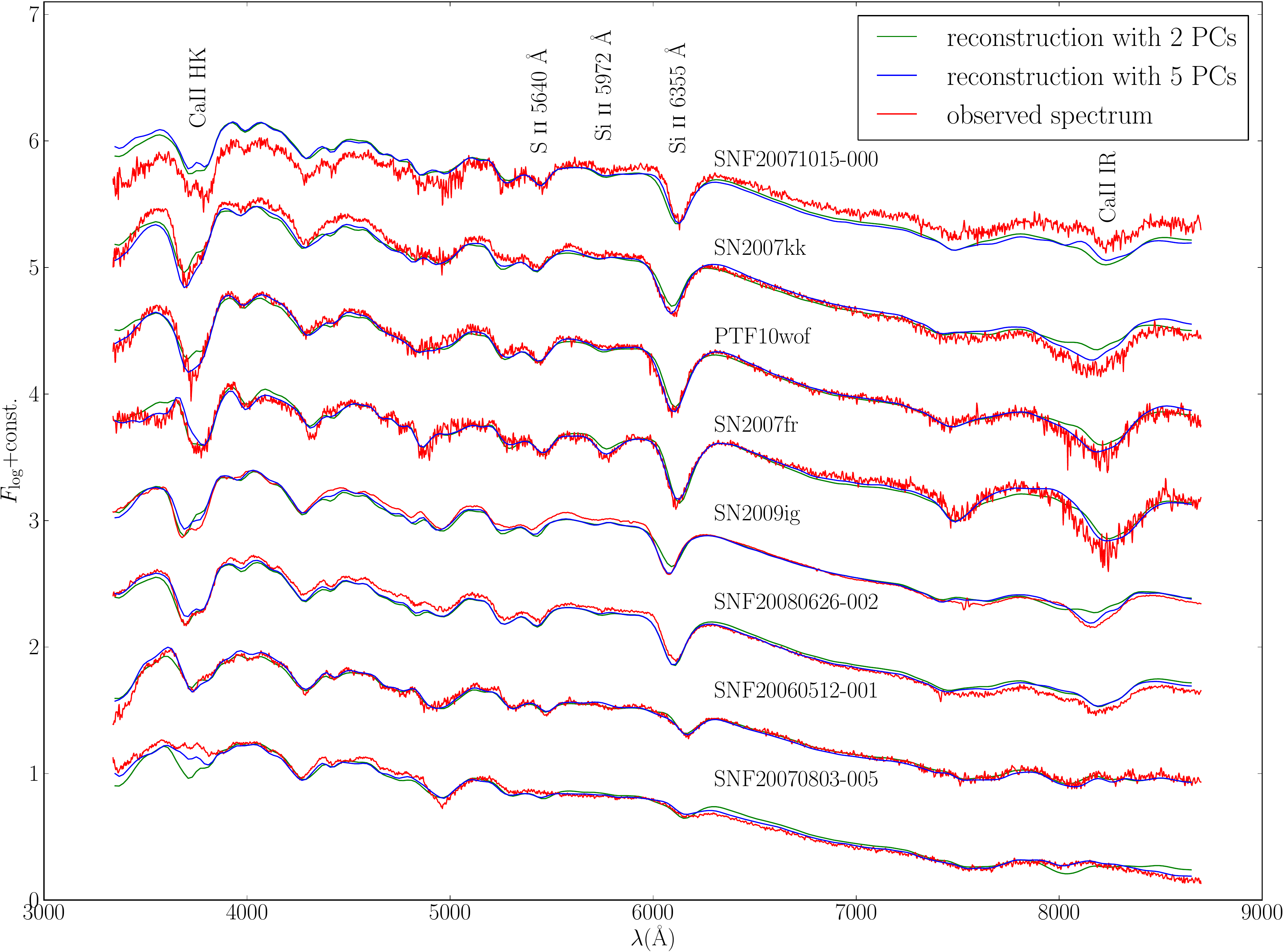}
\caption{Comparison between the observed spectra without smoothing (red) and reconstructed spectra using 2 (green) and 5(blue) PCs, in the $dF_{\log}$ approach, for a few supernovae at  $B-$band maximum light. 
}
\label{fig:rec_real}
\end{center}
\end{figure*}

In Figure \ref{fig:rec_real} some of the reconstructions are directly shown. We present the original spectra along with  reconstructed ones using two and five PCs.
 The plot shows a few SNe at maximum for clarity, but this behaviour holds for all epochs between $-10$ to $+10$ days. Here, the consequences of our choices in focusing on intrinsic features are obvious. Although the overall spectral shape and most lines are very well recovered, the ratio of fluxes at long wavelengths (color) is not. This is welcome and expected because the derivative analysis does not give much weight to the mean slope of the spectra, making the analysis independent of individual SNe reddening and reshapes the observed spectra so to allow a fair comparison with synthetic models.
The comparison in the derivative space is shown in Appendix \ref{sec:appendix_der_space}.

\subsection{Comparison with models}

\begin{figure}
\begin{center}
\subfloat{\includegraphics[width=1.\columnwidth]{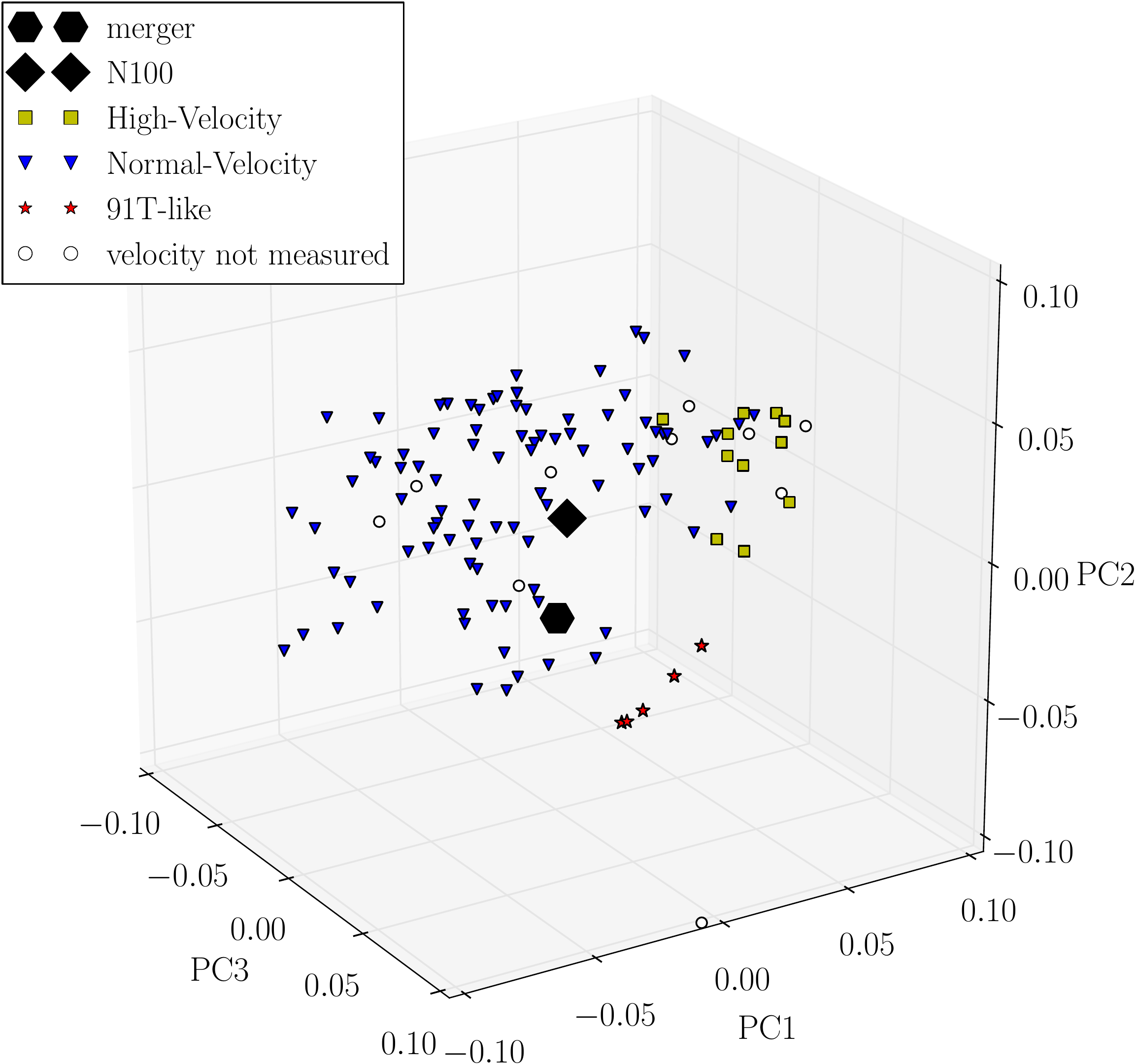}}
\caption{The merger (star) and the N100 delayed detonation (diamond) models, projected in to the first 3 components in derivative space.}
\label{fig:models}
\end{center}
\end{figure}

We emphasis the potential of this metric space, built from the combination of the derivative approach and EMPCA, 
in providing an ideal environment for the characterization of synthetic {spectral series (from $-10$ to $+10$~days)} within a space defined by observed SNe. As an example, the black symbols in Figure~\ref{fig:models} denote the projection, in derivative PC space, of the 3D delayed detonation model ``N100''  \citep{2013MNRAS.429.1156S,2013MNRAS.436..333S} and the merger model from \cite{2012ApJ...747L..10P}.

The N100 model describes a supernova generated from a white dwarf accreting material from a companion and getting close to the Chandrasekar mass (1.4 $\Msun$). The other model describes the explosion of two carbon-oxygen white dwarfs with masses of $0.9\Msun$ and $1.1\Msun$ prototypical for the double degenerate scenario.
 The model spectra were constructed based on  
 radiative transfer calculations of \citet{2009MNRAS.398.1809K} and projected into the PC space through the same procedure applied to the observed sample. 
Both models have been proposed as explanation of normal SN~Ia, in particular SN2011fe \citep[e.g.][]{2012ApJ...750L..19R}, and have a similar luminosity.

Figure~\ref{fig:models} demonstrates that our procedure does place both models among the normal SNe (blue dots), in derivative PC space. 
However, the geometrical distance between them is considerable, reflecting the intrinsic and spectral differences of these two models \citep{2012ApJ...750L..19R}. This illustrates the power of our method when applied to characterize SN~Ia models, providing an automatic and quantitative approach to confront them with observations and with each other. Such investigation  will be further developed in a subsequent work.

\section{Comparison with discrete observables}
\label{subsec:obs}

In the context of the PLS we will now study the correlation between the PC space and a few other photometric and spectroscopic quantities.
We present a closer look at each of these characteristics and describe in more detail how to obtain such information from the derivative PC space.

The absolute B-band magnitude at maximum is probably the most important quantity for the characterization of SNe~Ia. SNe~Ia are standardizable candles because a high degree of homogeneity in SN~Ia absolute magnitudes can be achieved using simple transformations based on parameters of their light curves. Given the crucial role played by these objects in astronomy and cosmology, a handful of techniques have already been developed aimed at properly standardizing them. The empirical relation between brightness and decline rate demonstrated by (Phillips 1993) is considered one of the first standardization techniques for SNe~Ia. It is given in terms of $\Delta$m15(B), which represents the decrease in B-band magnitude at 15 days after maximum brightness. Brighter SNe tend to decline more slowly and consequently present a lower value for $\Delta$m15(B). This standardization was substantially improved by introducing corrections based on broadband colors \citep{1996ApJ...473..588R,1998A&A...331..815T,1999AJ....118.1766P}. Ostensibly such color corrections account for extinction from dust, but most likely also contain a hidden color-luminosity correlation intrinsic to the SNe~Ia themselves.

For the purpose of comparing models with observations, any successful model should obtain the correct SN~Ia absolute magnitudes, and contain the brighter-broader relation. However, in the derivative PCA space the overall flux scaling and broad-wavelength color have been removed, and therefore are not directly represented in the derivative PCA space. Fortunately there are a number of spectroscopic indicators known to correlate with overall lightcurve peak brightness, width, and color. For instance, \cite{1995ApJ...455L.147N} found that the ratio between the depths of the \SiII~5972~\AA\ and the \SiII~6355~\AA\ lines correlates with peak B-band absolute magnitude. The pseudo Equivalent Width (pEW) at B-maximum of the \SiII~4000~\AA\ line correlates very well with lightcurve width \citep{2008A&A...492..535A,2008A&A...477..717B,2011AA...529L...4C}, as does that of \SiII~5972~\AA\ \citep{2006MNRAS.370..299H}. There is also evidence that the velocity of the \SiII~6355~\AA\ line is correlated with the intrinsic SN color \citep{2011ApJ...729...55F}.
 Since information related to pseudo equivalent widths and velocities will exist, and possibly be enhanced, by taking the flux derivative with respect to wavelength, it is quite likely that the derivative PCA space will retain the ability to differentiate between supernovae, and models, having different luminosities, lightcurve widths, and intrinsic colors. Here we apply PLS to explore the presence  of such correlations in our derivative PC space.

\begin{figure}
\begin{center}
\includegraphics[width=1.\columnwidth]{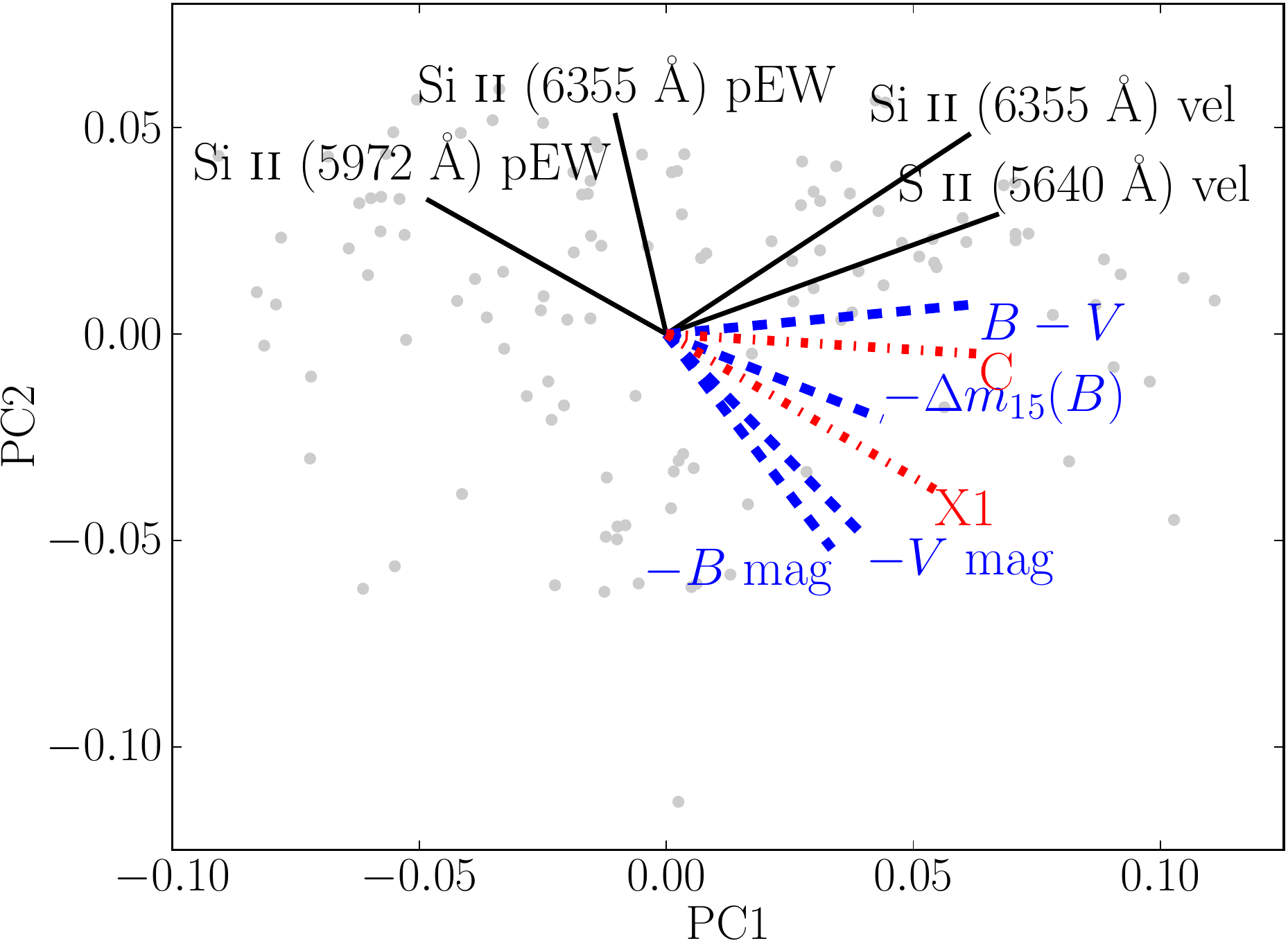}
\caption{The directions maximizing the covariance with various SN  
        parameters derived in a 5-dimensional space  and projected into the plane formed by
        the first two principal components.  Gray points are the same as those shown in Figure~\ref{fig:PCs_derivative}.  Directions correlated with spectroscopic quantities 
        are coloured in black (solid), photometric quantities in blue (dashed),
        and results from the SALT2 fit in red (dash-dotted). 
}
\label{fig:PCs_correlations}
\end{center}
\end{figure}

\

\subsection{Measurement of Discrete Observables}
\label{sec:res_pls}

We wish to measure the B-band magnitude at maximum and $\Delta$m15(B) with the fewest possible modeling assumptions. Wherefore, we simply fit a third order polynomial to the B magnitudes measured between $-10$ and $+25$ days from maximum {using errors coming from the noise of the spectra}. The fit is evaluated at maximum and at $+15$ days after maximum to obtain the peak $B$-band magnitude and $\Delta$m15(B), respectively. Uncertainties come from an error propagation of parameters from the polynomial fit. The $V$-magnitude at the epoch of $B$-band maximum is recovered from an analogous fit run on the $V$-magnitudes. The difference of the two magnitudes {at $B-$maximum} gives us the $B-V$ color. The input magnitudes are synthesized from our spectrophotometric time series, using the $B$ and $V$ filter responses given by \citet{1995PASP..107..672B}. Absolute magnitudes considered here are obtained from the observed apparent magnitudes at $B$-band maximum assuming Hubble-flow distances, without any extinction corrections. The errors on the absolute magnitudes are computed from the uncertainties in the light-curve fits and added in quadrature to uncertainties due to peculiar velocity of the host galaxies of $\sim 500$~km~s$^{-1}$ (Hawkins et al. 2003).

\begin{table}
\setlength{\tabcolsep}{4pt}
\renewcommand{\arraystretch}{1.5}
\begin{tabular}{lrrrrr}

\hline
 & PC1 & PC2  & PC3    & PC4 & PC5 \\
\hline
 D$\left( {\rm Si}~{\textrm{\sc ii \,}} 6355 {- \rm vel }\right)$ & $0.74$ & $0.58$ & $0.13$ & $0.24$ & $0.21$ \\ 
 D$\left( {\rm S}~{\textrm{\sc ii \,}}  5640 {- \rm vel }\right)$ & $0.81$ & $0.35$ & $0.35$ & $-0.21$ & $0.24$ \\ 
 D$\left( {\rm Si}~{\textrm{\sc ii \,}} 5972 {- \rm pEW }\right)$ & $-0.58$ & $0.39$ & $0.21$ & $0.33$ & $0.60$ \\ 
 D$\left( {\rm Si}~{\textrm{\sc ii \,}} 6355 {- \rm pEW }\right)$ & $-0.12$ & $0.64$ & $-0.38$ & $0.59$ & $0.30$ \\ 
\hline
 D$\left(B \text{mag}\right)$ & $0.40$ & $-0.63$ & $0.49$ & $-0.32$ & $-0.32$  \\ 
 D$\left(V \text{mag}\right)$ & $0.49$ & $-0.60$ & $0.49$ & $-0.21$ & $-0.34$  \\ 
 D$\left(B-V\right)$ & $0.76$ & $0.09$ & $0.43$ & $0.45$ & $-0.13$  \\ 
 D$\left(\text{c}\right)$ & $0.76$ & $-0.06$ & $-0.24$ & $0.58$ & $0.11$  \\ 
 D$\left(\Delta m_{15}\right)$ & $-0.53$ & $0.25$ & $0.07$ & $0.39$ & $0.71$  \\ 
 D$\left({x_1}\right)$ & $0.66$ & $-0.45$ & $-0.05$ & $-0.26$ & $-0.54$  \\ 
\hline
\end{tabular}
\caption{Directions in PC space found by PLS. Each direction is defined as a linear combination of the first 5 PCs whose coefficients are shown above (e.g., D$\left( {\rm Si}~{\textrm{\sc ii \,}} 6355 {- \rm vel }\right)$ = $0.74\times$PC1 + $0.58\times$PC2 + $0.13\times$PC3 + $0.24\times$PC4 + $0.21\times$PC5). }
\label{tab:PLS_directions}
\end{table}

As a point of comparison, we also performed light curve fits using the well-known Spectral Adaptive Lightcurve Template, (SALT2; Guy et al. 2007) code. SALT2 employs an internal model constructed using a linear PCA approach. The model is described by stretch ($x_1$) and color ($c$) parameters. The $x_1$ parameter is analogous to $\Delta$m15(B), while $c$ is analogous to $B-V$. Here the fits use magnitudes synthesized in the $BVR$ top-hat filters described in \cite{2013A&A...554A..27P}.

Here we focus on three key spectroscopic features: \SiII~6355~\AA,  \SiII~5972~\AA,  and \SII~5640~\AA. Technical details of the algorithm used to measure their spectroscopic pseudo equivalent widths and velocities directly from the SNfactory spectra are presented in Appendix \ref{ap:vel_pEW}. Since we do not possess a spectrum at maximum for all of our SNfactory SNe, we determined velocities and pEWs for every available spectra within $-7$ days and $+7$ days, for each SN. These values were then used to perform a linear fit from which we derived the values at maximum and corresponding uncertainties. We required a minimum of three successful measurements in this time window for the SN to be considered for the fit. This method proved to be quite robust, however, it is not capable of distinguishing the HVFs from the normal photospheric component, when both are present. Thus, every time we mention independently measured spectroscopic features, we are referring to the velocity of a given line, and not its HVFs counterparts.

\begin{table}
\centering
\renewcommand{\arraystretch}{1.5}
\begin{tabular}{lrl}
\hline
  & Pearson coeff. & $\sigma_{\textrm{res}}$    \\
\hline
 $ {\rm Si}~{\textrm{\sc ii \,}} 6355 {- \rm vel }$ & $0.85$ & $612$  \kms \\ 
 $ {\rm S}~{\textrm{\sc ii \,}}  5640 {- \rm vel }$ & $0.93$ & $351$ \kms \\ 
 $ {\rm Si}~{\textrm{\sc ii \,}} 5972 {- \rm pEW }$ & $0.85$ & $4.9$ \AA\ \\ 
 $ {\rm Si}~{\textrm{\sc ii \,}} 6355 {- \rm pEW }$ & $0.92$ & $9.9$ \AA\ \\ 
\hline
 $\Delta m_{15}$ & $0.78$ & $0.13$  \\ 
 ${x_1}$ & $0.74$ & $0.60$  \\ 
\hline
\end{tabular}
\caption{Pearson correlation coefficient for the  linear fit  between the directions found by PLS and independently measured observables. $\sigma_{\textrm{res}}$ corresponds to {the mean residual} between the measured observables values and those determined through PLS.}
\label{tab:PLS_sigma}
\end{table}

\subsection{Results from PLS}

{In Section~3.2, we saw that five components are sufficient to address most of the variance in the spectral features present in SNfactory data. Therefore, from now on we will work in a 5D PC space and use PLS to establish connections between these PCs and other independently measured parameters. Our goal is to demonstrate the potential encompassed by our derivative PC space, which summarizes the evolution of spectral features of a large SN~Ia sample. Using the nomenclature of in Section~2.5, the PLS technique was used to find the direction in 5D PC space ($\mathcal{X}$) which best describes each one of the SNe features cited in Section~4.3 (1D - $\mathcal{Y}$)}.

{Figure~\ref{fig:PCs_correlations} shows PLS results for the spectroscopic and photometric features discussed in Section~5.1, projected  -- for pedagogical reasons -- onto the first 2 PCs. Each one of these lines is obtained from a linear combination of the first 5 PCs, whose coefficients are presented in {Table~\ref{tab:PLS_directions}}. In this plot we see the first evidence of important physical information present in the derivative PC space: the connection between the pEW of \SiII~5972~\AA and $\Delta$m15(B).
As expected from the studies of \citet{1995ApJ...455L.147N} and \citet{2006MNRAS.370..299H}, the direction found
by PLS for the pEW of this line is similar to that of $\Delta$m15(B) (i.e. opposite to $-\Delta$m15(B), Figure~\ref{fig:PCs_correlations}). The velocity of \SiII~6355~\AA\ is seen to correlate with color, as expected from the study of \citet{2011ApJ...729...55F}. Interestingly, we also find a strong correlation of the velocity of \SII~5640~\AA\ with color. In terms of our PCs, we
find that PC1 correlates the best with indicators of color}.

{In Table~\ref{tab:PLS_sigma} we present the correlations given by PLS for SNe features with each one of the directions highlighted in Figure~\ref{fig:PCs_correlations}. The fact that many important SN features have strong signatures in our new metric spaces gives us confidence that our framework can help us better place synthetic spectra among their real data counterparts. Next we examine these trends in more detail}.

\begin{figure*}
 \begin{minipage}{.47\textwidth}
\begin{center}
\includegraphics[width=1.\columnwidth]{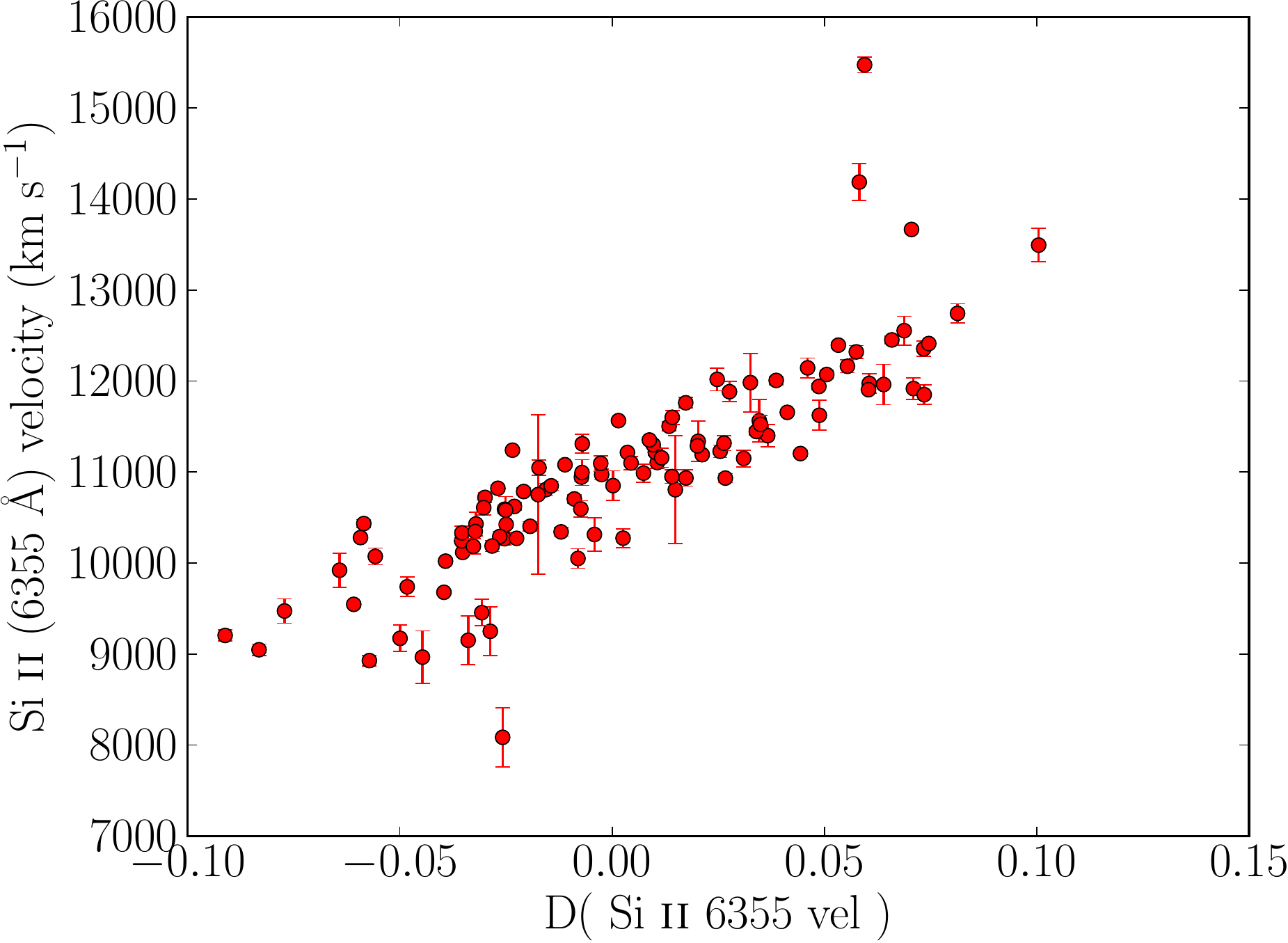}
\caption{Correlation between PLS result and the \SiII~6355~\AA\ velocity at $B$-band maximum. The few outliers on the high-velocity side are due to HVFs of Si (see text).}
\label{fig:SiII_vel}
\end{center}

 \end{minipage}
 \hspace{.04\linewidth}
 \begin{minipage}{.47\textwidth}

\begin{center}
\includegraphics[width=1.\columnwidth]{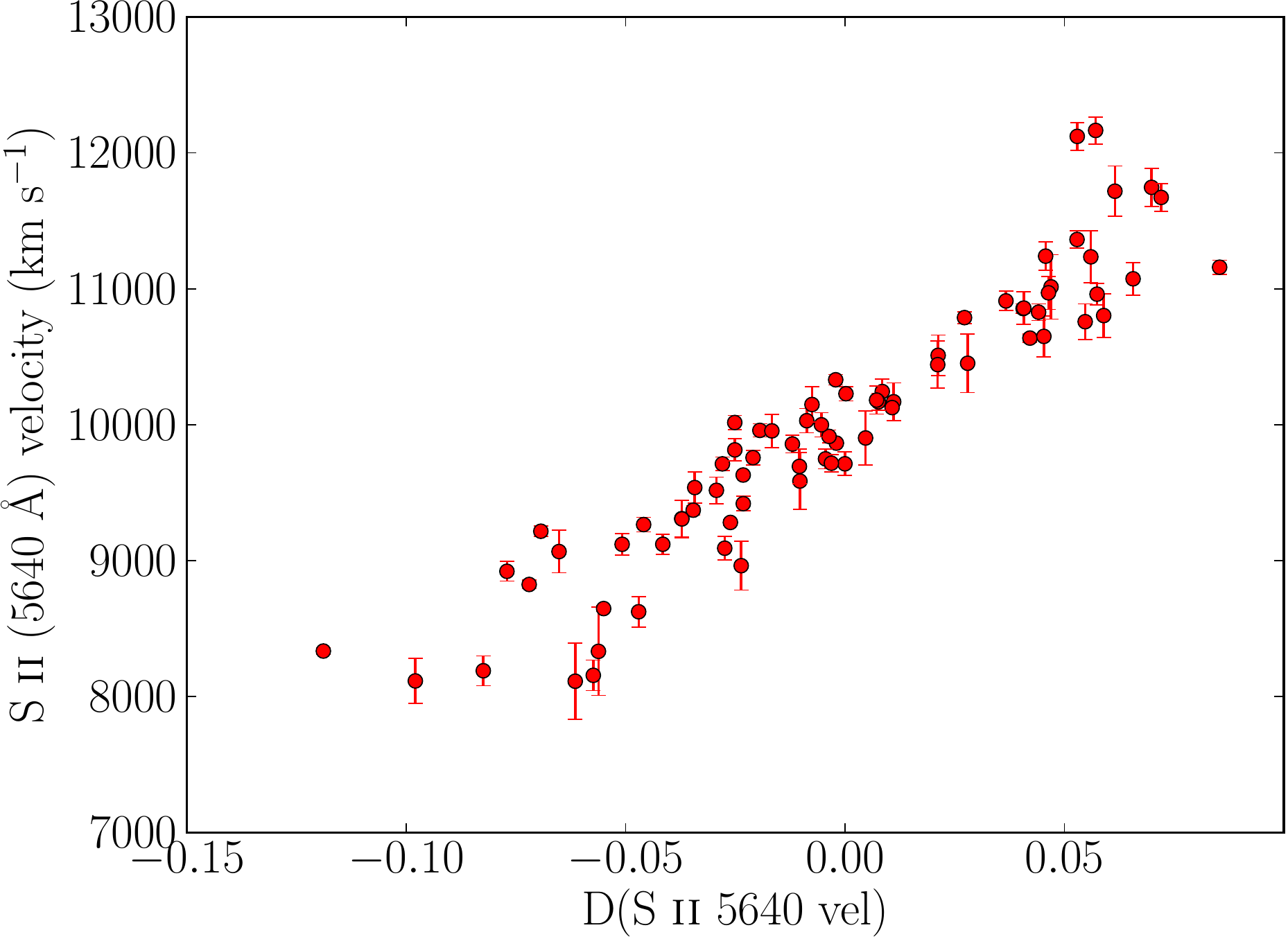}
\caption{Same as Figure \ref{fig:SiII_vel}, but for the \SII~5640~\AA\ velocity at $B$-band maximum. }
\label{fig:SII_vel}
\end{center}

 \end{minipage}
\end{figure*}

\begin{figure*}
 \begin{minipage}{.47\textwidth}
\begin{center}
\includegraphics[width=1.\columnwidth]{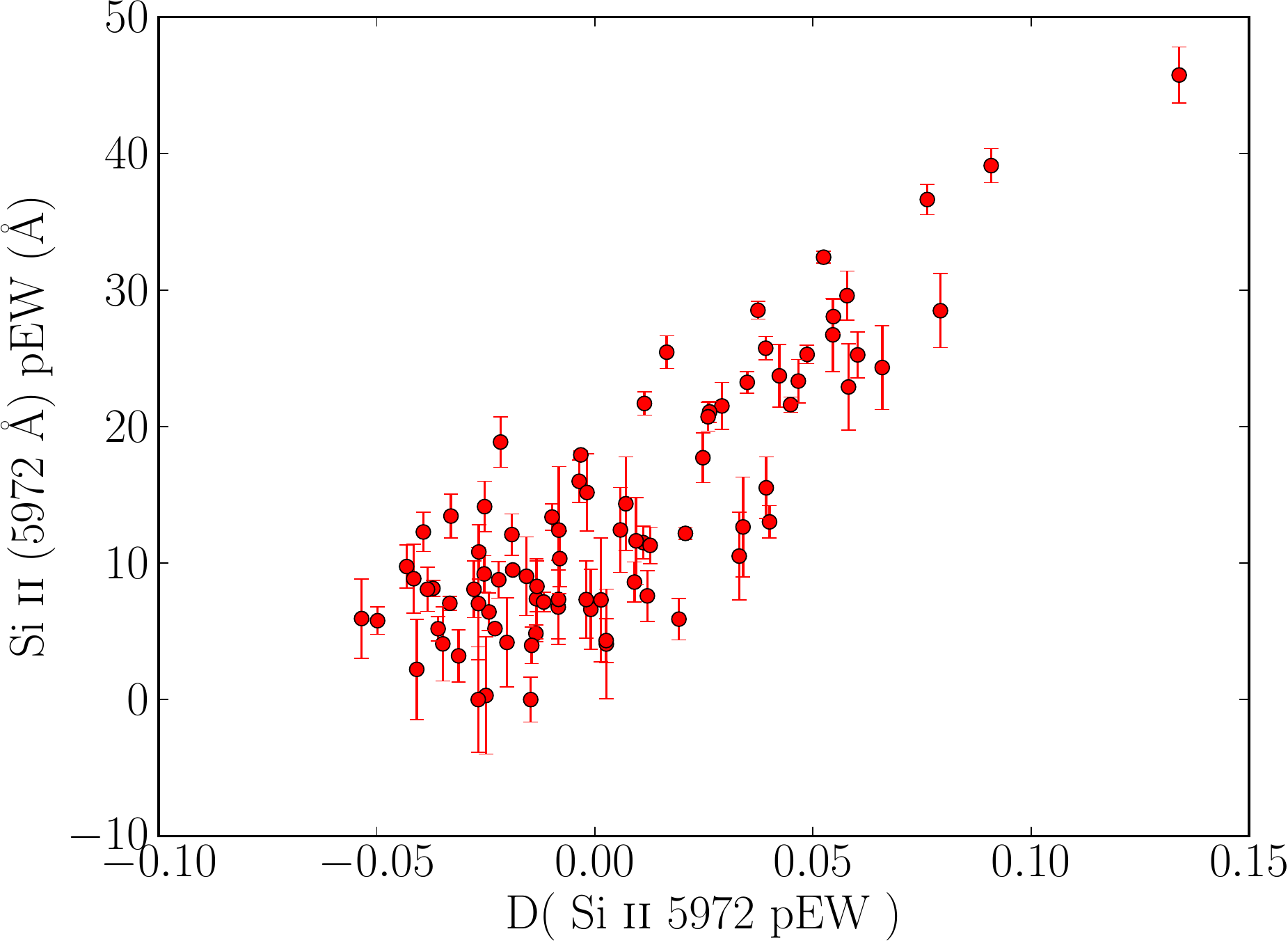}
\caption{Same as Figure \ref{fig:SiII_vel}, but for the \SiII~5972~\AA\ pEW at $B$-band maximum. }
\label{fig:SiII2_pew}
\end{center}

 \end{minipage}
 \hspace{.04\linewidth}
 \begin{minipage}{.47\textwidth}

\begin{center}
\includegraphics[width=1.\columnwidth]{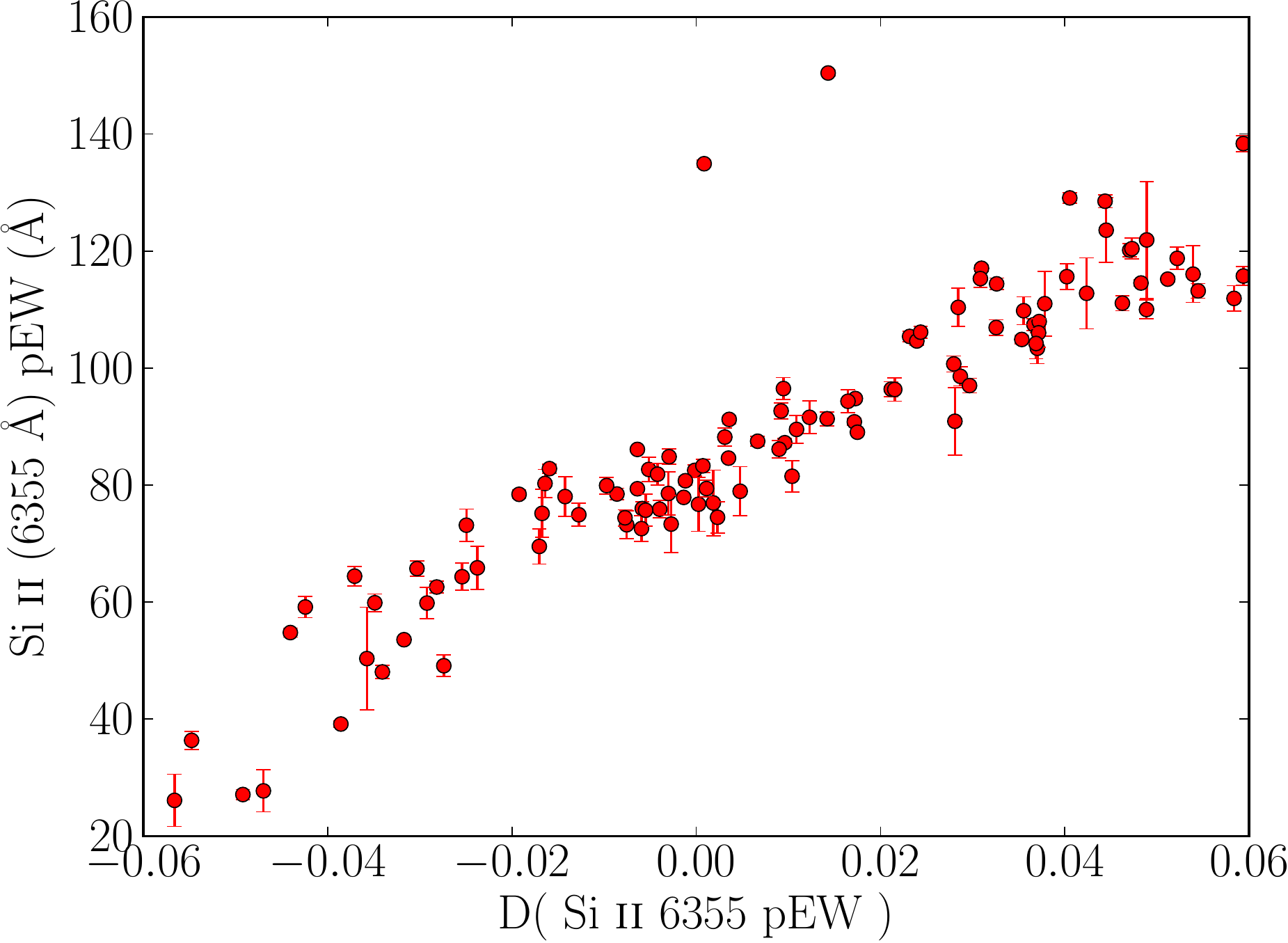}
\caption{Same as Figure \ref{fig:SiII_vel}, but for the \SiII~6355~\AA\ pEW. }
\label{fig:SiII_pew}
\end{center}

 \end{minipage}
\end{figure*}

\subsection{Spectroscopic observables in derivative PC space}

{Figure~\ref{fig:SiII_vel} shows the correlation between the velocity of \SiII~6355~\AA\ at maximum and the corresponding direction found by PLS in PC space. From {Table~\ref{tab:PLS_directions}}, we see that it is highly correlated with PC1 and PC2 but not so much with PC3, PC4 and PC5. This is still another angle on the HVFs discussed before: the velocity of \SiII\ is among the persistent features of SNe~Ia, and not correlated with the mechanism that gives rise to the HVFs of Ca lines (Section~4.1). The few outliers on the high-velocity side of Figure~\ref{fig:SiII_vel} are due to strong HVFs of \SiII\ still present around maximum. Their velocity is not predicted by the combination of components that predicts the photospheric velocity, suggesting also that the HVF of \SiII\ is not correlated with the main physics of the explosion and follows the more diverse behaviour of the outer layers}.

{Our ability to describe the velocity of \SII~5640~\AA\ using the 5D PC space is illustrated in Figure~\ref{fig:SII_vel}. Given the weakness of this line, the quality of the fit is quite impressive (Pearson correlation coefficient (PCC) is 0.93). This is not completely unexpected if one realizes that this line is usually narrower than the saturated \SiII~6355~\AA\ line, making possible a better measure of the velocity. More generally, \SII\ lines are not affected by HV features, which can complicate the measurement the photospheric component. These characteristics suggest that the velocity of \SII~5640~\AA\ might present a viable alternative to the \SiII~6355~\AA\ line for classification purposes. The \SII\ lines form deep in the ejecta and are good tracers of the photospheric velocity (Blondin et al. 2006). It is expected that for objects with similar luminosities and rise times, a larger photospheric velocity corresponds to a larger radius for the photosphere, a lower radiation temperature and, consequently, a redder color. The ability to extract such an effect from our derivative PC space is very promising as a tool for synthetic spectra characterization.
Finally, we emphasize that, although the PC space itself encompasses information regarding the entire time window study here ($-10$ to $+10$ days around B-band maximum), the directions obtained by PLS are bounded by the epoch in which the corresponding spectral features were measured. In this context, the correlations presented in Figs.~13 to 15 are only valid at maximum. An analogous study aimed at a different epoch would require the determination of spectral features at the epoch in question}.

{Figures \ref{fig:SiII2_pew} and \ref{fig:SiII_pew} show the correlation obtained by PLS between the pseudo equivalent widths of \SiII~5972~\AA\ and of \SiII~6355~\AA\, which are the basis of the \cite{2006PASP..118..560B} classification scheme. These have Pearson correlation coefficients of 0.85 and 0.92, respectively. This is another indication that information used by others to differentiate between SNe~Ia strongly persists in the derivative PCA space}.

\begin{figure*}
 \begin{minipage}{.47\textwidth}
\begin{center}
\includegraphics[width=1.\columnwidth]{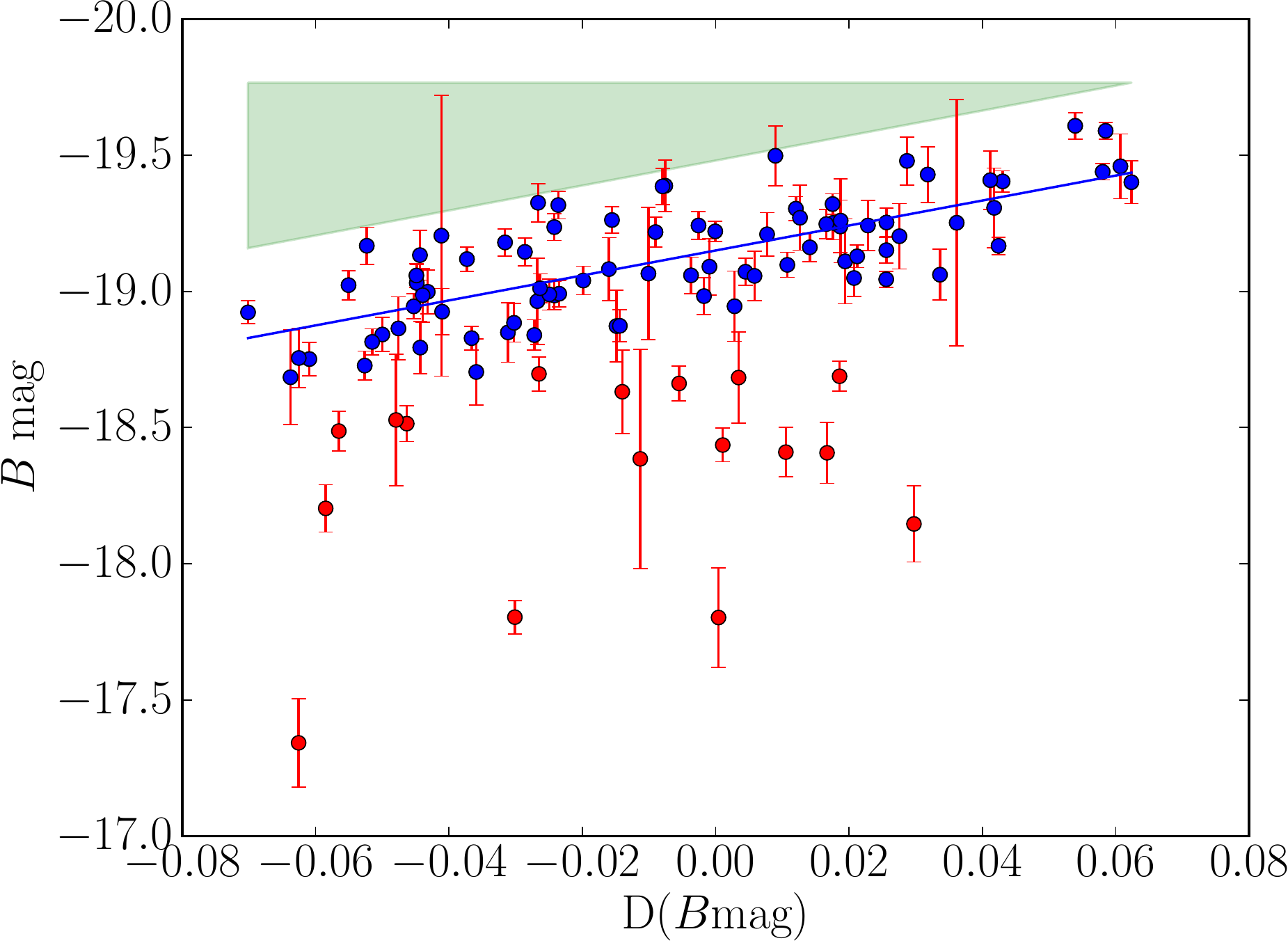}
\caption{Correlation between PLS result and the $B$-band magnitude. The red points belong to SNe much redder than others with the same spectral characteristics.
}
       \label{fig:Bmag_vs_PC}
\end{center}
 \end{minipage}
 \hspace{.04\linewidth}
 \begin{minipage}{.47\textwidth}
\begin{center}
\includegraphics[width=1.\columnwidth]{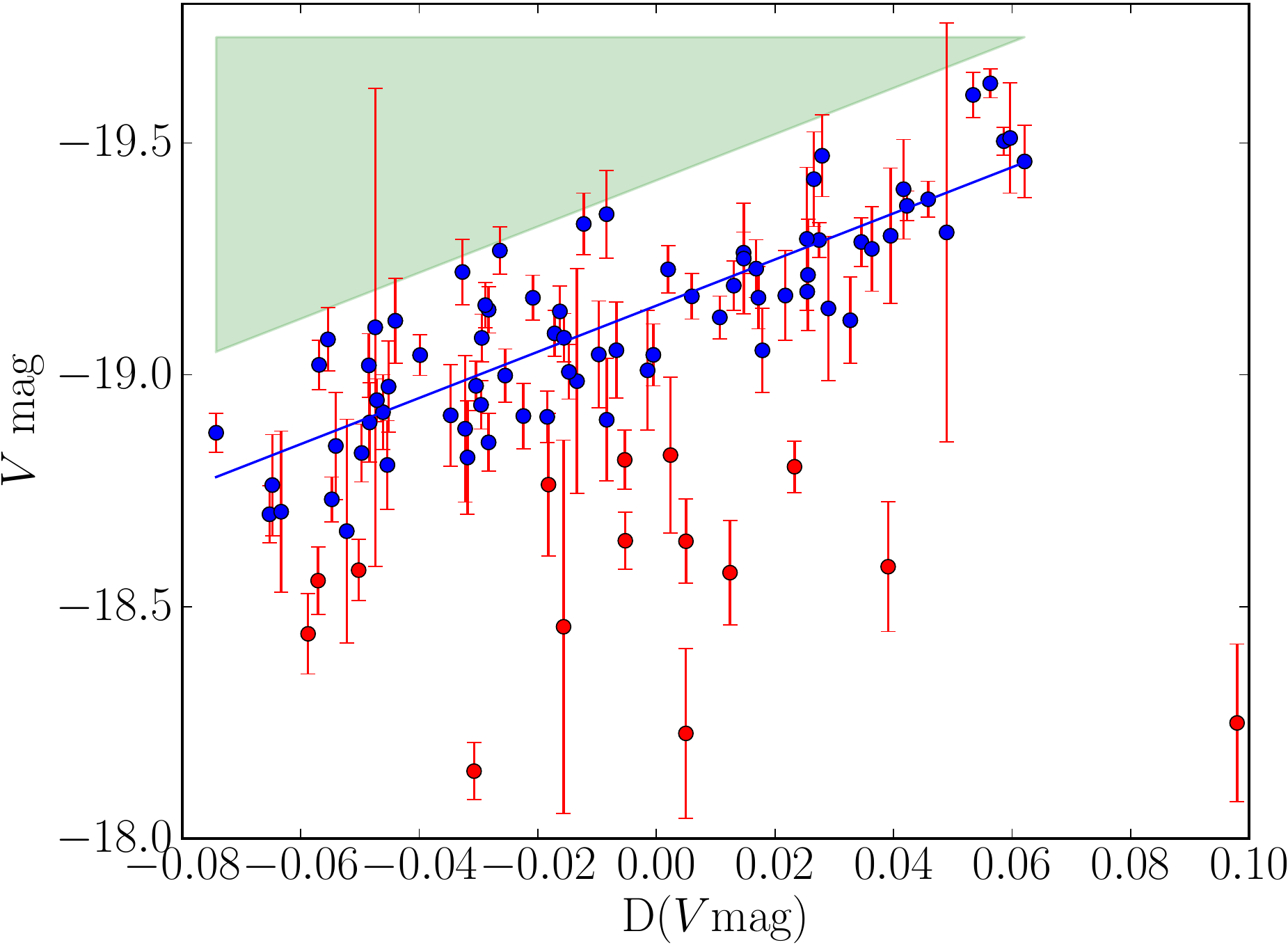}
\caption{Same as Figure \ref{fig:Bmag_vs_PC} for $V$-band magnitude.}
\label{fig:Vmag_vs_PC}
\end{center}
 \end{minipage}
\end{figure*}

\begin{figure*}
 \begin{minipage}{.47\textwidth}
\begin{center}
\includegraphics[width=1.\columnwidth]{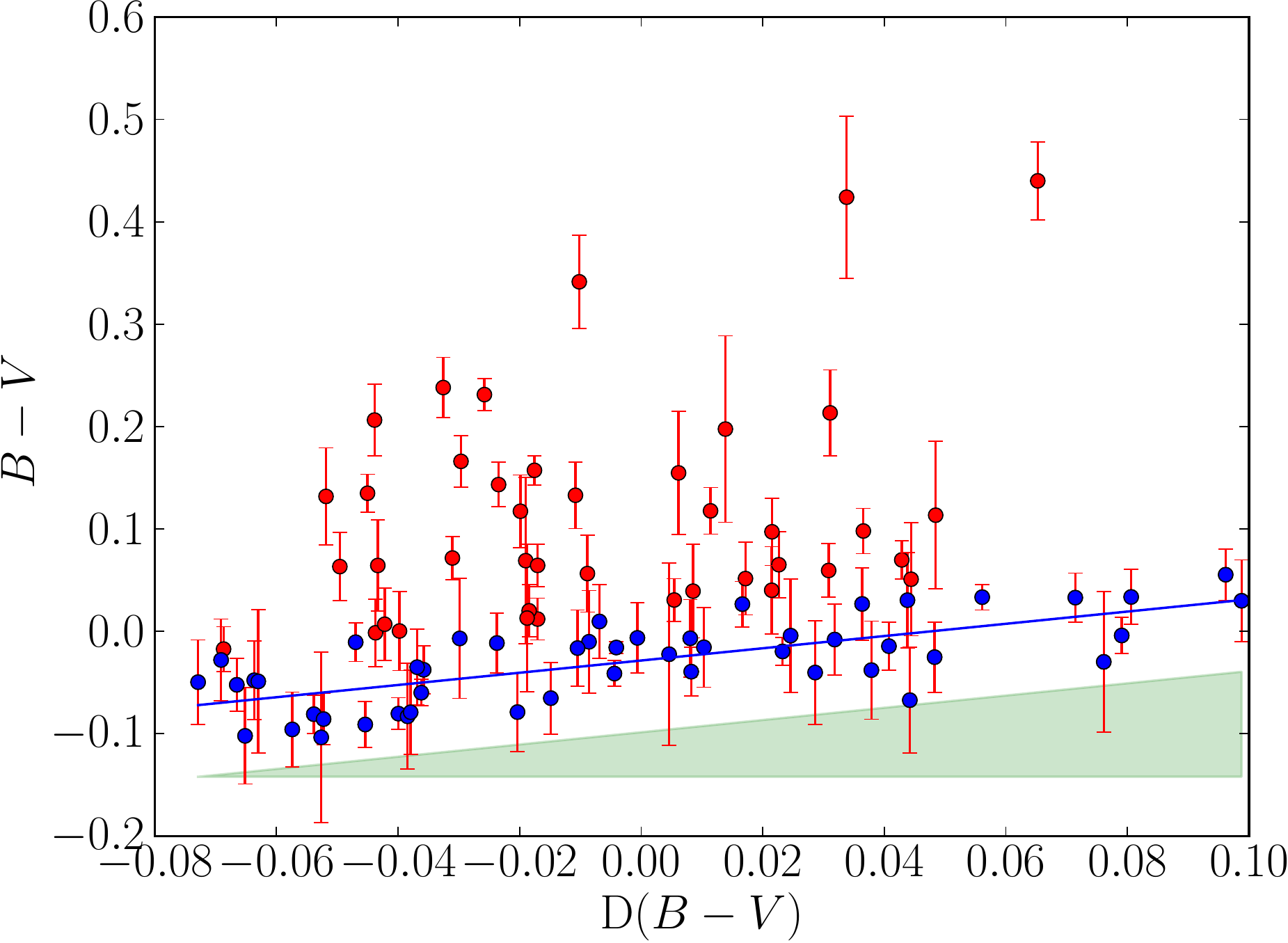}
\caption{Same as Figure \ref{fig:Bmag_vs_PC} for the $B-V$ color. The red points belong to significantly reddened supernovae
         (E(B-V)$\gtrsim 0.1$), and the blue points represent almost 
         unreddened ones. 
}
\label{fig:BmV_vs_PC}
\end{center}

 \end{minipage}
 \hspace{.04\linewidth}
 \begin{minipage}{.47\textwidth}
\begin{center}
\includegraphics[width=1.\columnwidth]{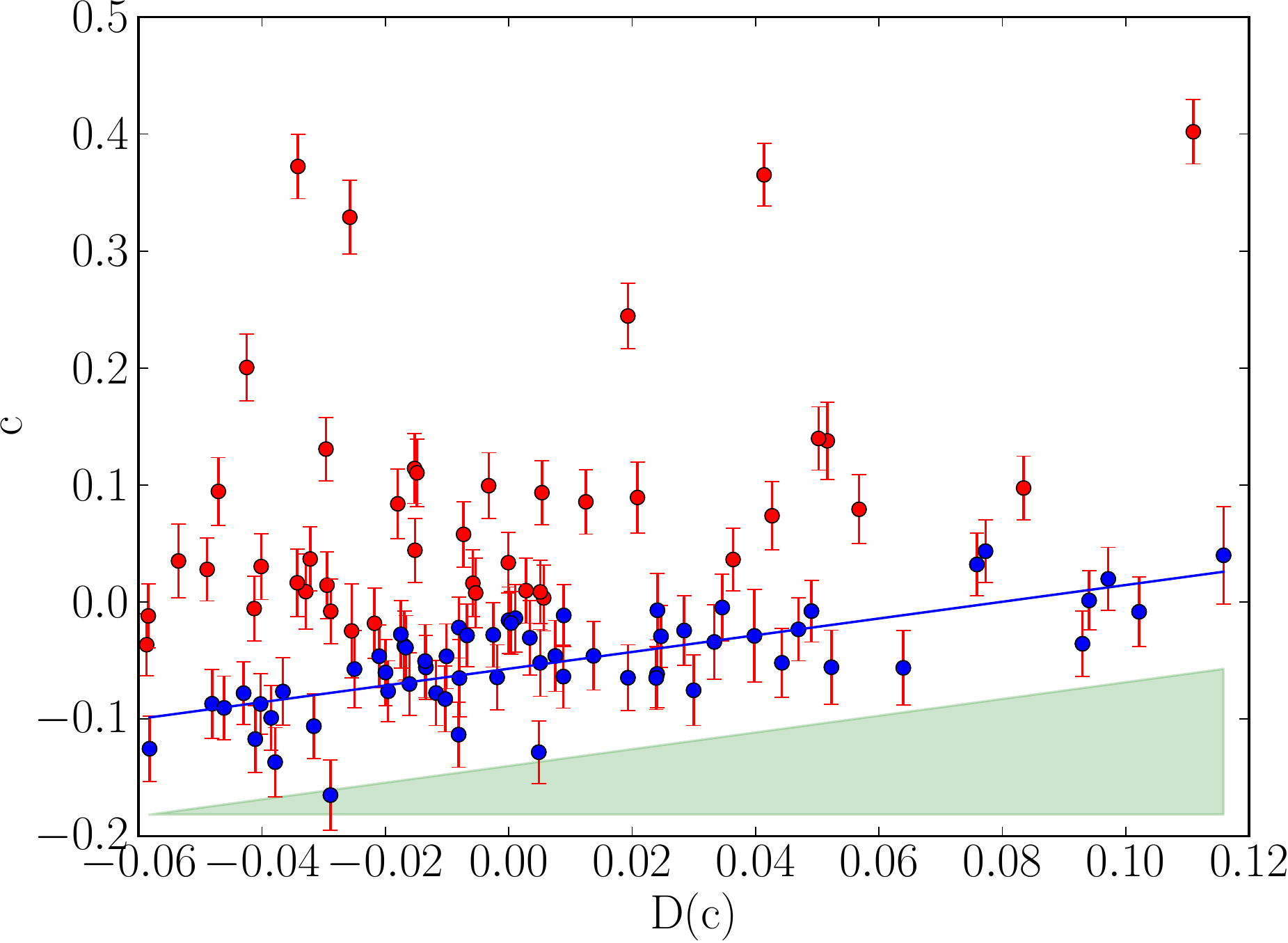}
\caption{Same as Figure \ref{fig:BmV_vs_PC} for SALT2 color parameter $c$.  }
\label{fig:color_salt}
\end{center}
 \end{minipage}
\end{figure*}

\begin{figure*}
 \begin{minipage}{.47\textwidth}
\begin{center}
\includegraphics[width=1.\columnwidth]{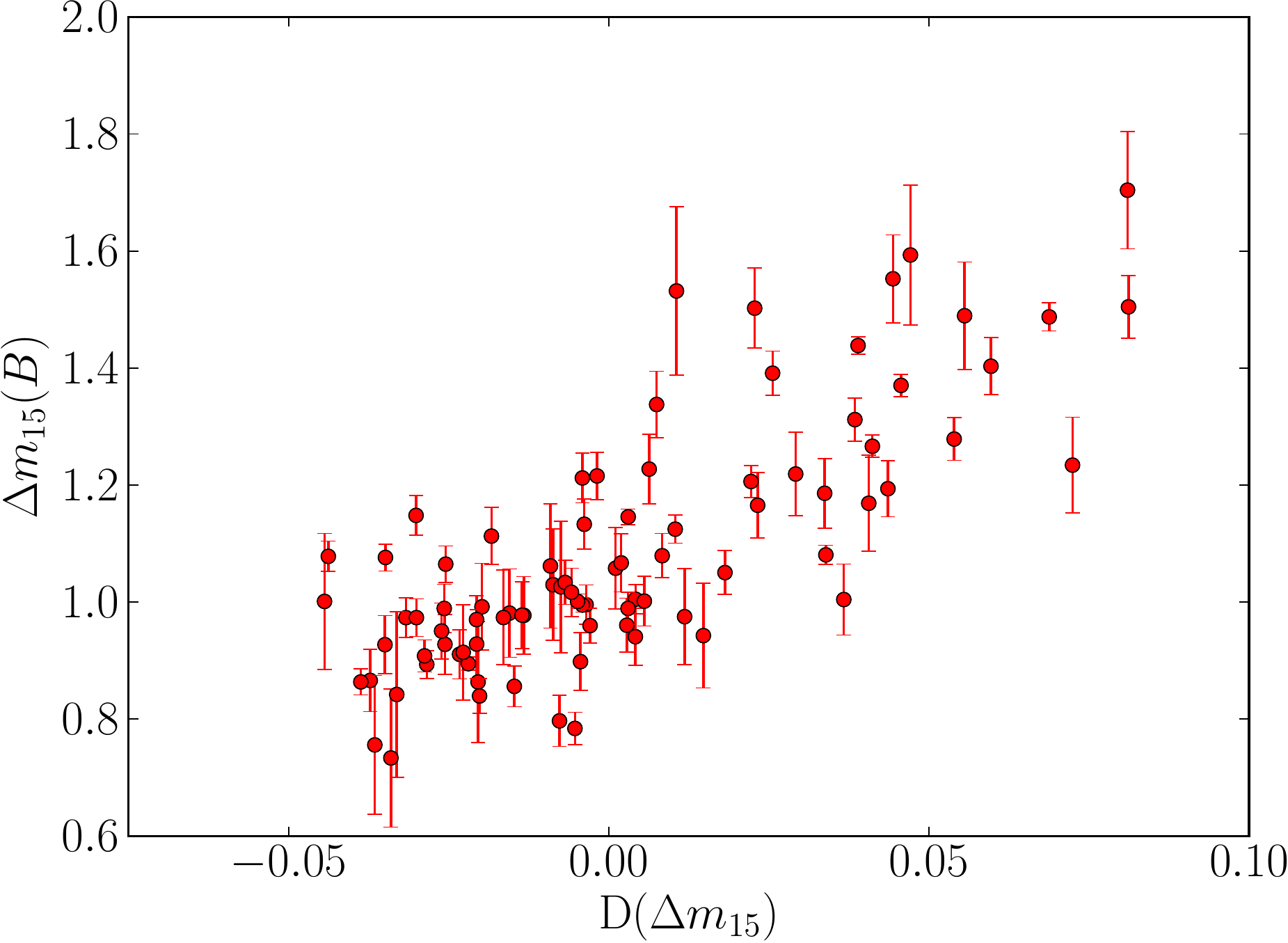}
\caption{Correlation between PLS result and \Deltam\ . The horizontal axis represents 
         direction in 5D PC space which most correlates with \Deltam\ and the vertical axis is the value for this parameter measured from the SNe light-curves.
        }
\label{fig:DM15_vs_PC}
\end{center}
 \end{minipage}
 \hspace{.04\linewidth}
 \begin{minipage}{.47\textwidth}
\begin{center}
\includegraphics[width=1.\columnwidth]{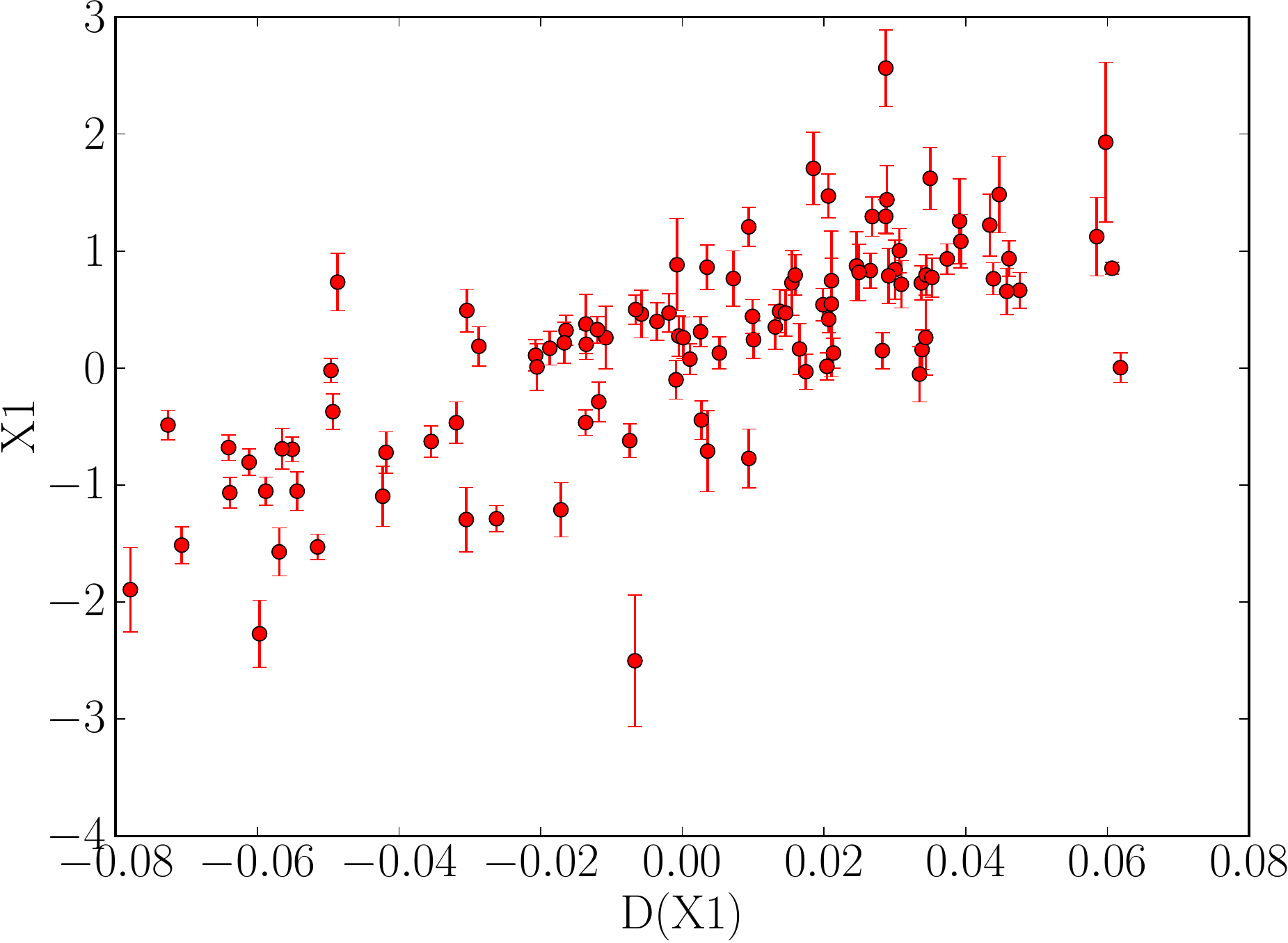}
\caption{Same as Figure \ref{fig:DM15_vs_PC}, but for the $x_1$ parameter of SALT2. \vspace{0.8cm}}
\label{fig:X1}
\end{center}
 \end{minipage}
\end{figure*}

\subsection{Photometric observables in derivative PC space}

{Having established correlations between spectroscopic luminosity and color indicators and our 5D PC space, we expect to find correlations with B-band peak magnitudes and colors. However, unlike the spectroscopic features discussed above, or the photometric $\Delta$m15(B) parameter, these are strongly affected by dust extinction and reddening. {The information on the amount of the extinction is not present in the $dF_{\log}$ space. This means that observed colors and magnitudes cannot be completely reconstructed using this technique alone.} Nonetheless, it is of interest to examine these dust-polluted parameters since their intrinsic behavior is critical for understanding SN~Ia physics and standardization for cosmology. This may also allow advances in separating the intrinsic and dust contributions.}

{Figures \ref{fig:Bmag_vs_PC} and \ref{fig:Vmag_vs_PC} show the correlation between the directions in PC space and the observed B and V absolute magnitudes, respectively. All points represent rest frame magnitudes corrected for Milky Way but not for host-galaxy reddening. The well defined upper envelope situated  below the green triangles in each plot suggests a locus potentially dominated by SNe~Ia with little extinction. The presence of a slope to this upper envelope versus D(B) and D(V) is likely due to SNe~Ia suffering little dust extinction. Because D(B) and D(V) are largely free of the effects of extinction, this strongly suggests that the derivative PC space contains information on the intrinsic luminosity of SNe~Ia.} 

{In an effort to find the approximate direction of the luminosity vector, we attempt to isolate the least extincted SNe~Ia, under the assumption that brighter SNe~Ia have less extinction, using an iterative rejection scheme. This type of approach is common when attempting to establish intrinsic peak magnitudes for many SN~Ia standardization methods, however, it assumes that D(B) and D(V) impose a sufficient degree of order in the relative SN~Ia luminosities, which may be an oversimplification (e.g., \citet{2013A&A...560A..66R}).  (The crispness of the upper envelope is encouraging in this regard.) We applied PLS to the entire data set and then performed a linear fit between the observed magnitudes and the output direction in PC space. Based on this linear fit, only supernova brighter than the linear fit, or fainter by less than 0.3 mag, are selected to the next iteration.  PLS was applied again to the chosen subset and the process was repeated until convergence. The algorithm converged rapidly to a direction that represents the variation of the brightest SNe~Ia absolute magnitudes with D(B) or D(V). We found that the output direction in PC space depends only weakly on the criteria used to reject SNe in each iteration. Blue points in Figures~\ref{fig:Bmag_vs_PC} and \ref{fig:Vmag_vs_PC}  correspond to SNe selected in the final PLS iteration, the blue line denotes the final linear fit, and red points represent rejected objects.} 

{A similar procedure can be applied to color, using the assumption that the bluest SNe~Ia suffer the least amount of reddening by dust. This assumption is only effective if  D(B-V) imposes a sufficient amount of homogeneity in the SNe~Ia colors. Again, the crispness of the blue end of the color envelope offers encouragement that this is a sensible approach. Figure~\ref{fig:BmV_vs_PC} shows the correlation between $B-V$ color at maximum and the direction in PC space found by the iterative process described above. As in previous plots, each point corresponds to a color measurement without any attempt to correct for host galaxy reddening. The surviving SNe (blue dots) represent objects whose reddening is consistent with the locus of bluest objects to within their measurement errors. Figure~\ref{fig:color_salt} shows that the SALT2 $c$ parameter has a similar behaviour. This was expected from the existence of a correlation with $B-V$ color at B-band maximum, however the $c$ parameter incorporates the color information at other epochs included in the SALT2 fit. Hence, $c$ corresponds to a more general measurement of the SN~Ia color. Here again, because D(B-V) and D(c) are largely free of the effects of extinction, this strongly suggests that the derivative PC space contains information on the intrinsic color of SNe~Ia.} 

{Finally, Figure~\ref{fig:DM15_vs_PC} illustrates the correlation between $\Delta$m15(B) and the corresponding PLS result in PC space. The Pearson correlation coefficient between these two quantities is 0.78 (Tab. 2). Discrepancies frequently come from a wrong estimation of the decline rate. Comparing
the polynomial fit used to compute the $\Delta$m15(B) with the SALT2 $x_1$, the later usually gives better results. The SALT2 fit takes into account all epochs in B, V and R bands, obtaining a decline rate parameter quite consistent with the one suggested by the EMPCA analysis (Figure~\ref{fig:X1}), for most of the objects. This is also reflected in the similar directions found to correlate with $\Delta$m15(B) and $x_1$ in Figure~\ref{fig:PCs_correlations}}.

{We emphasize that the correlations between directions in PC space and global photometric properties like $x_1$ and $\Delta$m15(B) represent yet another test of the information encompassed in the metric space.} As it was constructed from the entire spectral sequences, it was expected to reproduce such photometric observables even though they were not inserted as features directly in the data matrix. This reinforces our statement that important information
is preserved throughout the entire process.

\section{Conclusions}
\label{sec:conclusions}

We have developed a new framework {which allows the simultaneous characterization of} large samples of spectra, {forming an ideal ground for placing synthetic spectra among the observed ones}. Combining Expectation Maximization Principal Component Analysis (EMPCA) and Partial Least Square (PLS) techniques, it defines a meaningful metric space  and  correlates it to spectroscopic and photometric intrinsic properties. 

The algorithm is based on the derivative of the spectrum over wavelength, which consequently assigns a larger weight to small scale features and, at the same time,  makes results independent of distance measurements, reddening and spectra calibration. The method allows an automatic  exploration of information encoded in weak spectral features from the weak lines themselves, not only through their correlation with stronger lines.  Moreover, the initial data matrix was forged to encode spectral evolution information through the use of spectral sequences representing each object.
{This shows an easy way to use the available spectral evolution information.}

We applied the method to a large sample ($\sim 120$ SNe and $\sim 800$ spectra) of well observed type Ia supernovae  obtained by the SNfactory collaboration. At first, we defined a low dimensional parameter space using EMPCA and studied the spectral features covered by each PC separately. Results  
show that the High Velocity Features  (HVFs) of \CaII~H\&K and infrared lines are uncorrelated with properties of the rest of the ejecta, consistent with \cite{2005ApJ...623L..37M}. This suggests that the outer layers of the ejecta have variations partially unrelated to {the} inner structure.

{We confirmed many of the results of \cite{2011MNRAS.410.2137C}. For example, the 91T-like SNe form a continuum of properties with normal SNe, PCA can be used to form a continuum of spectral templates, and the first two PCs mainly describe spectral velocities and equivalent widths. A larger dataset and the innovative method of analysing the derivative of the spectra allows us to have a stable metric space without arbitrarily removing peculiar objects from the sample.    }

Once the PC space was defined, we applied the PLS algorithm in order to find directions in this low dimensional space which correlate with independently measured SNe~Ia characteristics. In other words, we used the PC space as a tool which enables the reconstruction of not only the observed spectra, but also as a substitute  of the spectral parameters most used to sub-classify SNe~Ia: velocity and pseudo-Equivalent Width (pEW) of \SiII~5640~\AA\ and \SiII~6355~\AA\ lines, $B$  and $V$ magnitudes and $B-V$ color at maximum, \Deltam\ and SALT2 parameters $c$ and $x_1$.
This demonstrates that the PC space is physically meaningful and includes the information recovered from usual spectral indicators. Moreover, it clarifies the potential of this framework in finding missing or unexpected features in synthetic spectra.

Our PLS results confirm the well known correlation between the pEW of \SiII~5972~\AA\ and the \Deltam\ in SNe~Ia \citep{2006MNRAS.370..299H,1995ApJ...455L.147N}.
{The technique is not optimized to calibrate SN~Ia. The observed color and magnitudes cannot be directly reconstructed by this technique alone, because they are largely contaminated by extinction.    
}
 We show that the intrinsic $B-V$ color of SNe~Ia is not constant among different objects and correlates with the velocity of \SiII~6355~\AA, as found by \cite{2011ApJ...729...55F}.
We showed that the velocity of the \SII~5640~\AA\ can be used for the same scope. 

Now that we have the PCA trained on a large enough sample, this tool can be applied for direct comparison between synthetic and real SN~Ia spectra. Projecting a synthetic spectral series in this PC space will reveal its counterparts among the real data, by an analysis of its neighbours. Moreover, the relations discovered by PLS can independently characterize each spectral feature of a given model and place it according to its most important physical properties within the real data parameter space.

{Given the challenge of performing a coherent statistical comparison between synthetic and real spectra, our method can also be used to characterize complete sets of models built with different explosion scenarios. It is able to provide important insights regarding  the global properties of each explosion mechanism in order to favour or disfavour them. Such a global analysis should be more robust against systematics in the models than comparing them with individual SNe.
From a large enough synthetic spectra library, the method also allows the construction of a  PC space based entirely on models and the projection of real objects in it, providing a cross-check between the real and synthetic metric spaces. A detailed study of such applications will be investigated in a future work.}

\section*{Acknowledgments} 

We thank all the python, numpy and scipy communities for the high-quality free software they made available.
MS thanks Philipp Edelmann for all the technical support during the developing of this work.
 EEOI acknowledges financial support from Brazilian agencies FAPESP (2011/09525-3) and CAPES (9229-13-2).
Supported by German DFG Cluster of Excellence ``Origin and Structure of the Universe'' and the DFG Transregio Project 33 ``Dark Universe''.
We thank Dan Birchall for observing assistance, the technical and scientific staffs of the Palomar Observatory, the High Performance Wireless Radio Network (HPWREN), and the University of Hawaii~2.2 m telescope.
 We recognize the significant cultural role of Mauna Kea within the indigenous Hawaiian community, and we appreciate the opportunity to conduct observations from this revered site.
 This work was supported by the Director, Office of Science, Office of High Energy Physics, of the U.S. Department of Energy under Contract No. DE- AC02-05CH11231; by a grant from the Gordon \& Betty Moore Foundation; in France by support from CNRS/IN2P3, CNRS/INSU, PNCG, and the Lyon Institute of Origins under grant ANR-10-LABX-66 ; and in Germany by the DFG through TRR33 "The Dark Universe." Some results were obtained using resources and support from the National Energy Research Scientific Computing Center, supported by the Director, Office of Science, Office of Advanced Scientific Computing Research, of the U.S. Department of Energy under Contract No. DE-AC02-05CH11231. HPWREN is funded by National Science Foundation Grant Number ANI-0087344, and the University of California, San Diego.
This work was written using the  collaborative ShareLaTeX platform.

\label{lastpage}

\appendix

\section{Cross-validation}
\label{ap:kfolding}

\begin{figure*}
\begin{center}
\includegraphics[width=2.0\columnwidth]{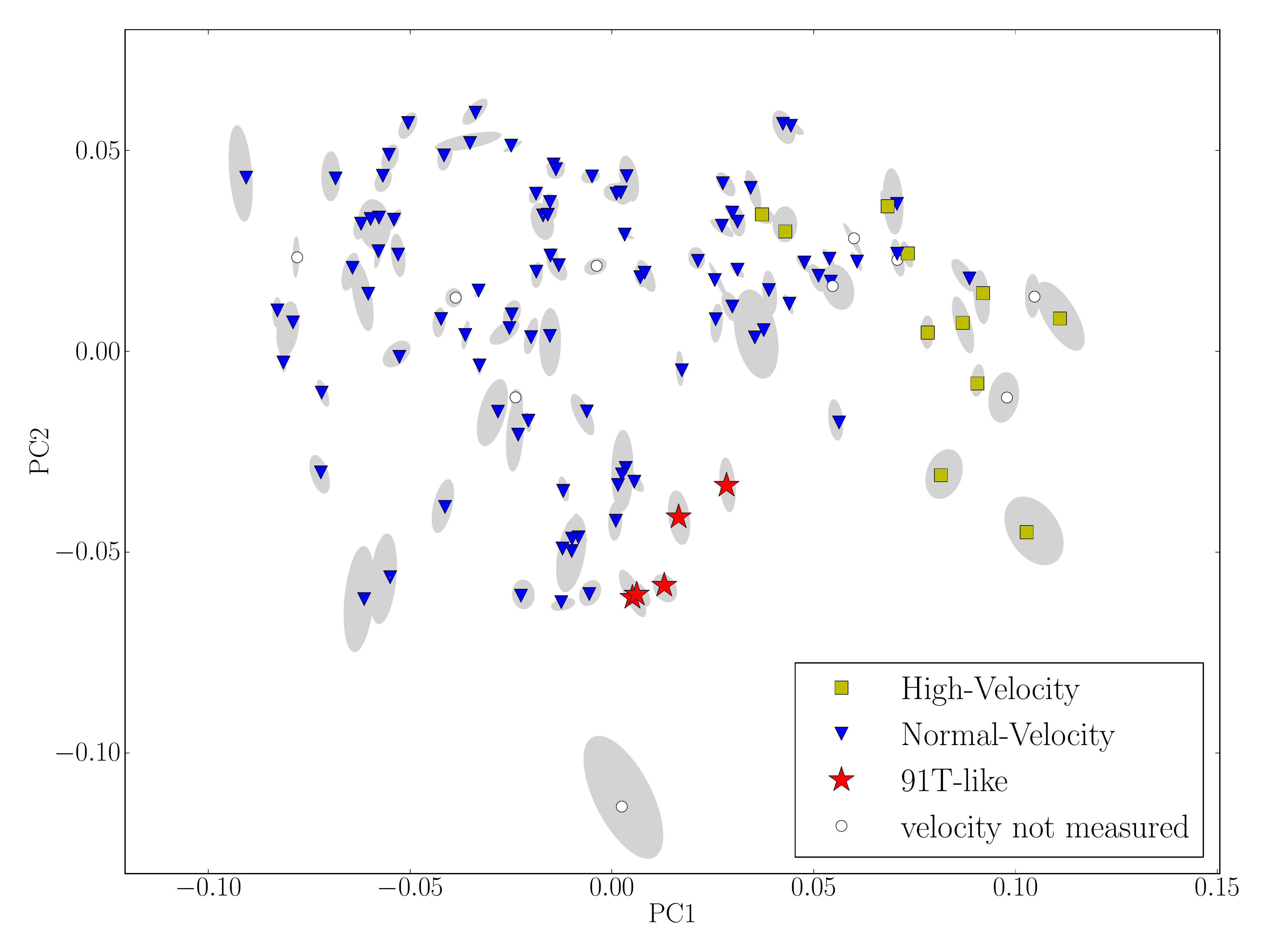}
\caption{Stability of PC space through $k=10$ folding cross-validation. The color code for points and line are the same used in Figure~\ref{fig:PCs_flux}. The gray ellipses denote mean and $1\sigma$ variance for locations occupied by each data point throughout the 10 iterations.}
\label{fig:PCs_deriv_ellipses}
\end{center}
\end{figure*}

\begin{figure*}
\begin{center}
\subfloat{\includegraphics[width=1.\columnwidth]{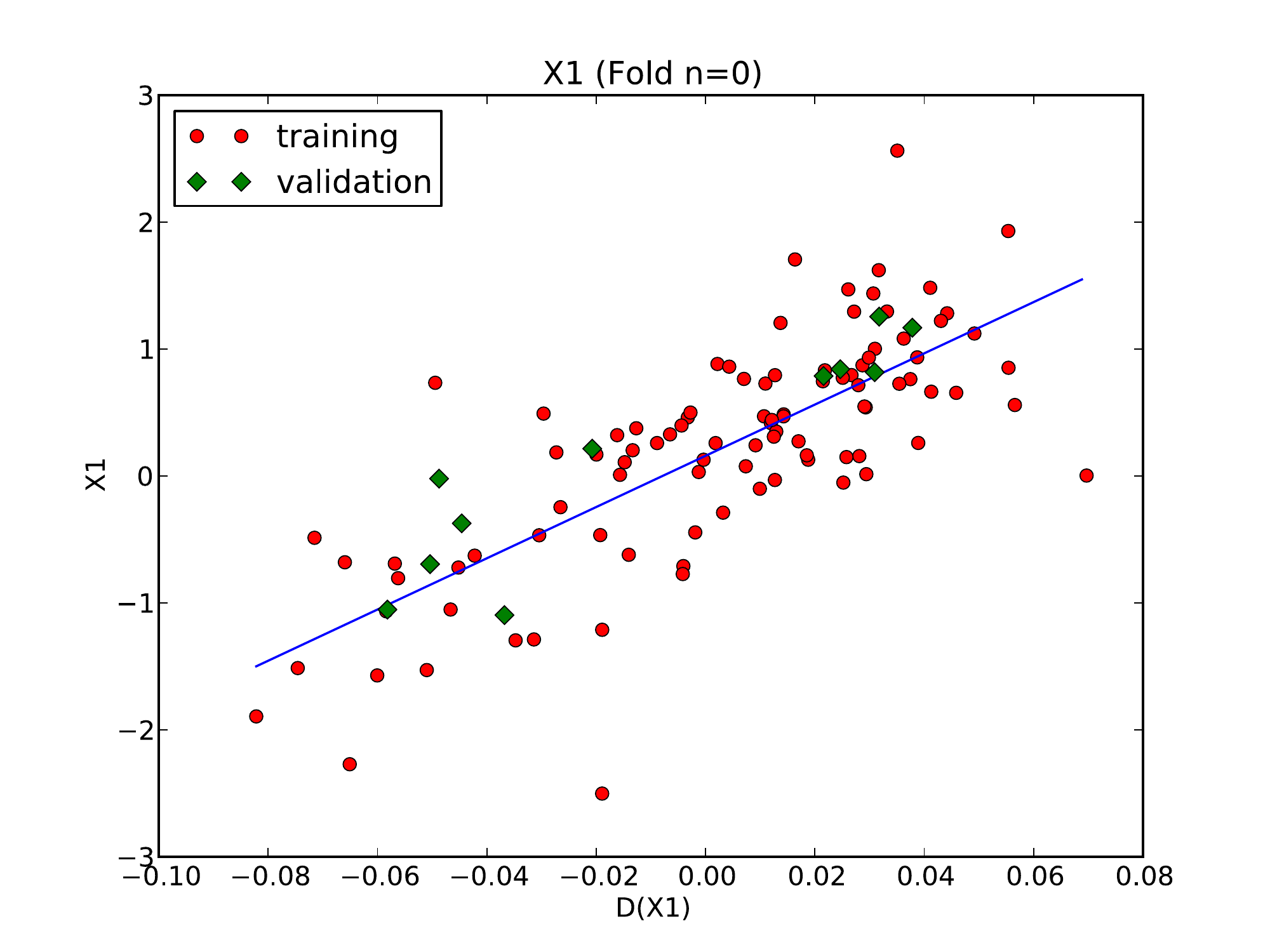}}
\subfloat{\includegraphics[width=1.\columnwidth]{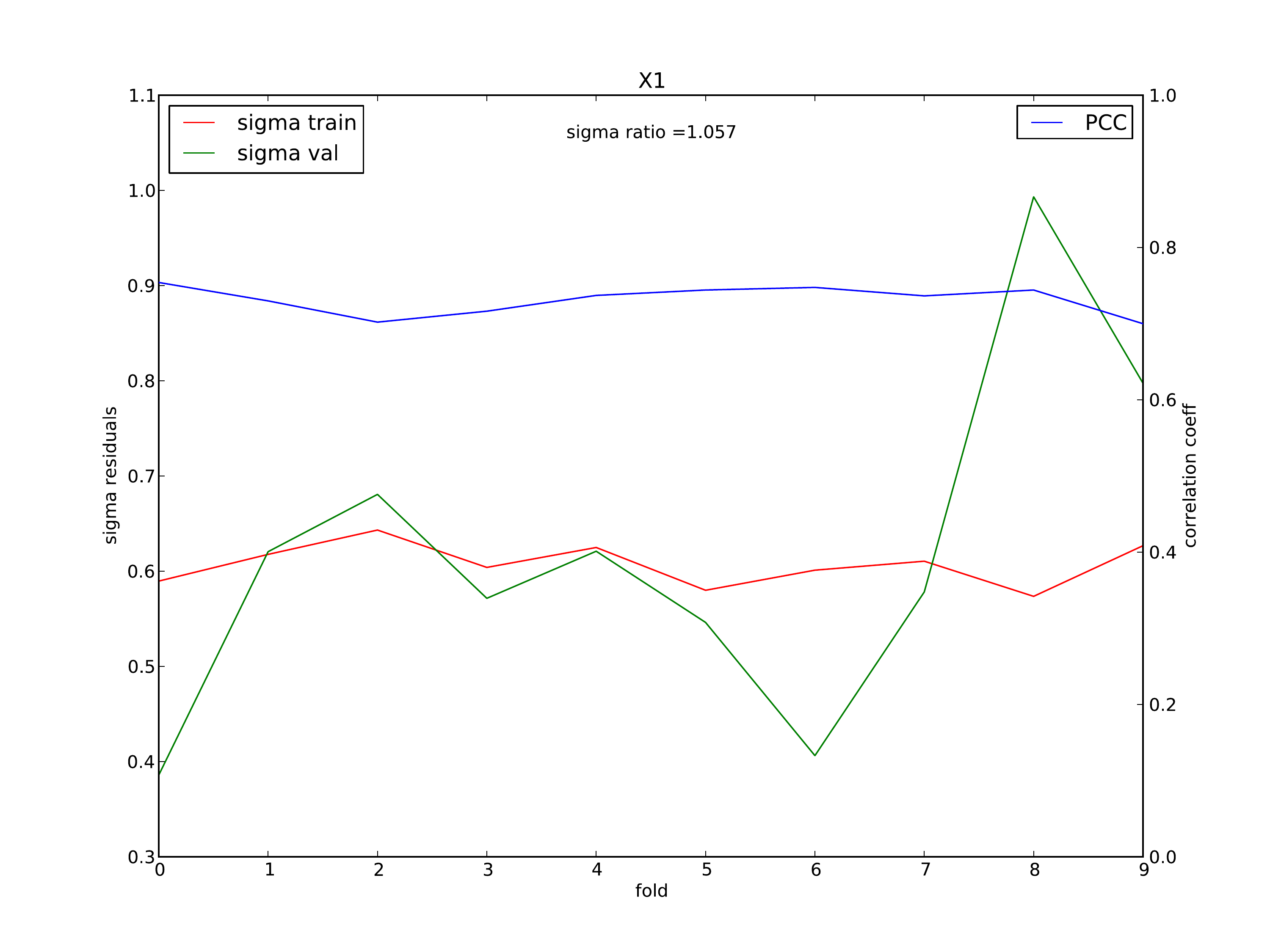}}
%
\caption{Accuracy of PLS analysis in predicting the value of $x_1$ for the validation sample. \textbf{Left panel:} Results for one of the realizations. The red circles and green diamonds correspond to the training {and} validation sets respectively. The blue line shows the result from the linear fit applied to the training sample only. \textbf{Right panel:} Residuals from training (red) and validation (green) samples shown on the left axis, and Pearson correlation coefficient (PCC, in blue), shown in the right axis, for all 10 iterations.  The average ratio between validation and training sample residuals is $\approx 1.057$.}
\label{fig:kfold_sigma}
\end{center}
\end{figure*}

We tested the stability of our PC space using a $k$-folding cross-validation (CV) algorithm. 
The goal of any CV procedure is to ensure that results are statistically consistent and not particular to a specific data set. At the same time, it tests for over-fitting. In our context, this means that even when applied to a sub-sample of the original data (training sample), the PC space configuration (Figure~\ref{fig:PCs_derivative}) must be recognizable. Moreover, the directions found by PLS in this space must be able to predict the values of the discrete observables for data not used in the EMPCA analysis (validation sample), using only their projections in PC space. Such results are expected to have residuals of the same magnitude for training and validation samples. 

The number of foldings ($k$) denotes how the data will be divided between training and validation samples. The original set is divided into $k$ mutually exclusive sub-samples and for each iteration one of these is stripped out of the original data set. The complete EMPCA and PLS algorithm is then applied to the remaining data and a linear fit is obtained characterizing the directions found by PLS and the discrete observables analysed in section \ref{subsec:obs}. This process is repeated for all $k$ subsamples and results for the PC projection and PLS analysis are stored in each iteration. The average displacement of each point in the PC space, calculated over all iterations,  gives us a measurement of how much the stability of this space relies on individual data points. If the PC space configuration is highly unstable for different subsets, it can be considered evidence of the need of a larger, more representative, sample in order to safely draw conclusions. Analogously, an over-fitting method can be recognized  if the  PLS analysis is not able to provide estimations of the discrete observables for objects in the test sample, at least as accurately as it does for the training sample. 

Here we present results for  $k=10$ foldings, which is a standard first choice for many CV procedures \citep{2009arXiv0907.4728A}. However, we did perform the test for different values of $k$, with results following the expected behaviour: the PC space becomes more stable  for larger values of $k$, the linear fits on the PLS results remain the same and the ratio of residuals between training and validation sets remain close to unity (Table~\ref{tab:PLS_sigma_kfold}).

The stability of the PC space in $dF_{\log}$ is shown in Figures~\ref{fig:PCs_deriv_ellipses} and \ref{fig:kfold_sigma}.
 The color code is the same used in Figure~\ref{fig:PCs_flux}  and the gray ellipses represent the mean and $1\sigma$ variance of the locations occupied by each data point in all the 9 realizations in which it was part of the EMPCA.
 As an example, we show in the left panel of Figure~\ref{fig:kfold_sigma}, the PLS results regarding the determination of $x_1$, in one of the iterations.
 This plot illustrates  how well  the PLS is able to determine values of $x_1$ for points in the validation sample (green diamonds) in comparison with the variance present in the training sample (red circles).
 A more quantitative approach to such results throughout all the CV process is shown in the right panel of the same figure.
 Residuals from the determination of $x_1$ for training (red) and validation (green) samples, as well as the Pearson correlation coefficient (blue) for different folds ($k$) are shown.
 The mean ratio between residuals from validation and test samples was found to be very close to unity, verifying that our method is not suffering from over-fitting in the determination of discrete observables. Similar tests were performed for other observables and numerical results are shown in Table \ref{tab:PLS_sigma_kfold}.

\begin{table}
\renewcommand{\arraystretch}{1.5}
\begin{tabular}{lllc}
\hline
& $\sigma_{\textrm{res}}$ & $\sigma_{\textrm{res}}$  & \multirow{2}{*}{ratio}\\
& training & validation & \\
\hline
$ {\rm Si}~{\textrm{\sc ii \,}} 6355 {- \rm vel }$ & $608$ \kms & $642$ \kms & $1.06$ \\
$ {\rm S}~{\textrm{\sc ii \,}}  5640 {- \rm vel }$ & $348$ \kms & $362$ \kms & $1.04$ \\
$ {\rm Si}~{\textrm{\sc ii \,}} 5972 {- \rm pEW }$ & $5.5$\AA  & $6.0$\AA  & $1.09$ \\ 
$ {\rm Si}~{\textrm{\sc ii \,}} 6355 {- \rm pEW }$ & $10.3$\AA  & $11.0$\AA  & $1.07$ \\
\hline
$\Delta m_{15}$ & $0.13$ & $0.15$ & $1.10$  \\
${x_1}$ & $0.61$ & $0.64$ & $1.06$  \\ 
\hline
\end{tabular}
\caption{Residuals in estimation of observables from training and validation samples.}
\label{tab:PLS_sigma_kfold}
\end{table}

\section{Line velocities and pseudo-EW calculations}
\label{ap:vel_pEW}

The values for line velocities and pEW used in section \ref{subsec:obs} were calculated using the algorithms described below.

In order to calculate the velocity of a line known to exist at an observed wavelength $\lambda_0$, we start by searching for local minimum around $\lambda_0$. Once the local minimum is found, we use its wavelength, the rest frame wavelength of the line  and add relativistic corrections to compute the velocity blueshift.  

If the line does not exist, the search for local minimum will lead us to the next important spectral feature and the final velocity value will be easy to recognize as wrong.

In computing the pEW, we need to determine the line tangent to the two nearest peaks surrounding a given spectral feature. We begin from the point of minimum flux of that feature (point A) and define two other points, along the flux function, to the left (point B) and to the right (point C) of point A. The area between the line connecting points B and C is calculated for successive small increments in the distances between A and B.  The algorithm continues to iterate until the area between line BC and the flux function stop increasing. Once this maximum area is reached, B is kept fixed and the same procedure is applied to successive small increments in the distance between A and C. The calculation continues to alternate between increments in AB and AC until convergence.
Once the maximum area is determined, it is used to characterize the pEW.

\section{The reconstructions in the derivative space}
\label{sec:appendix_der_space}

\begin{figure*}
\begin{center}
\includegraphics[width=2.0\columnwidth]{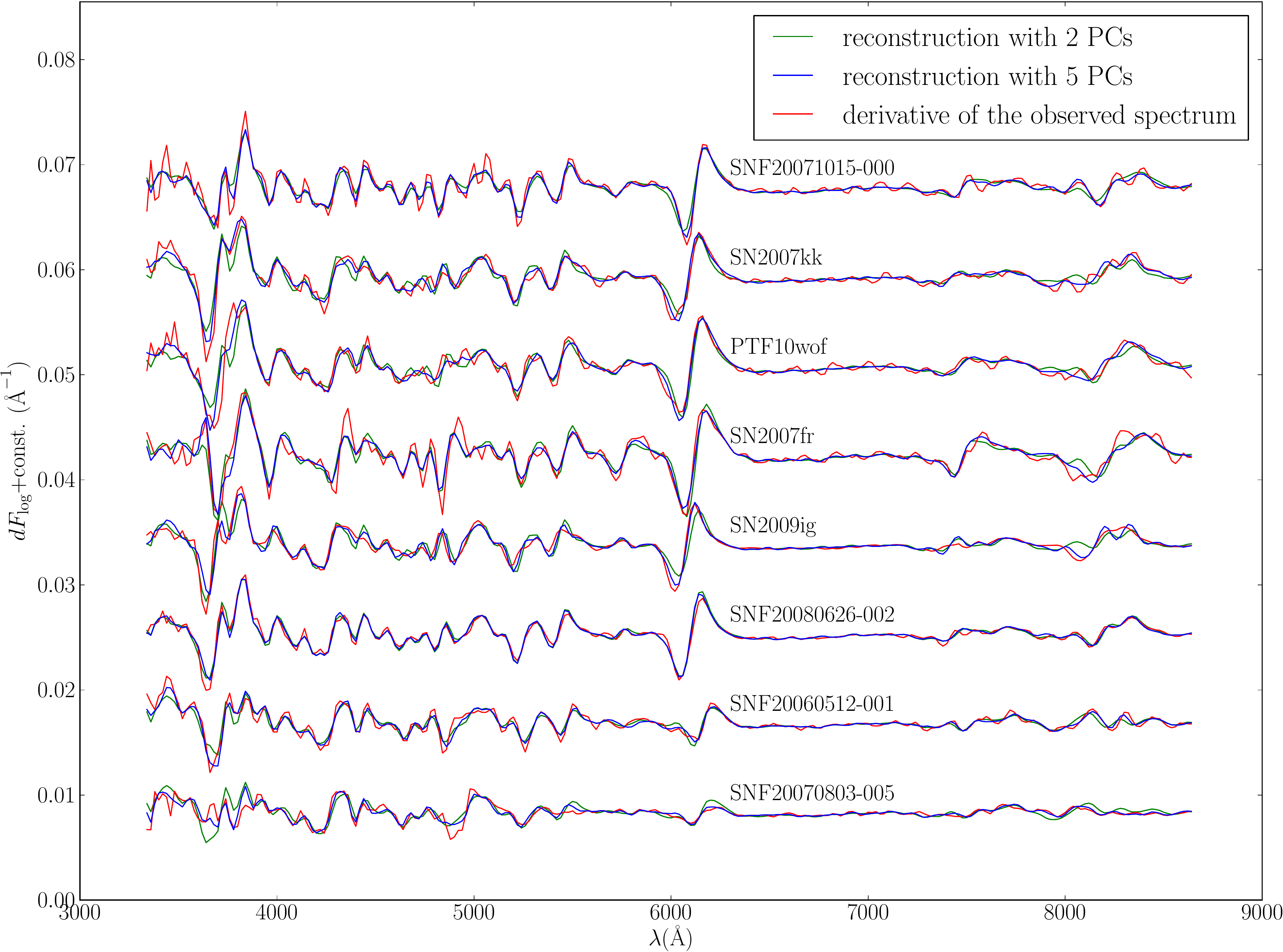}
\caption{Comparison between the derivative of the observed spectra (red) and reconstructions from PCA using 2 (green) and 5(blue) PCs for a few supernovae at $B-$band maximum light. 
}
\end{center}
\label{fig:rec_der}
\end{figure*}

In Figure~\ref{fig:rec_der} we show the same reconstructions presented in Figure~\ref{fig:rec_real} in the original derivative space.
We lack a physical intuition in observing this space and it is hard to recognise the behaviour of the classical spectral indicators.
However, it clearly demonstrates the ability of the derivative operation in minimizing reddening  effects. 
It is instructive that the mismatches in color, which appear in the first two objects in Figure~\ref{fig:rec_real} (SNF20071015-000 and SN2007kk), are not noticeable anymore.

%
%
%
 
\bibliography{biblio}


\end{document}